\documentclass[12pt]{article}
\usepackage{cite}
\usepackage{bm}
\usepackage[dvipdfmx]{graphicx}
\usepackage{amssymb}
\usepackage{mathtools}
\usepackage{epstopdf}
\usepackage{verbatim}
\usepackage{color}
\usepackage{slashed}
\usepackage{bbold}
\usepackage{float}
\usepackage[title]{appendix}
\usepackage[export]{adjustbox}
\usepackage{cancel}
\usepackage[colorlinks, citecolor=blue, linkcolor=red]{hyperref}
\usepackage{amsfonts}
\usepackage{amssymb}
\usepackage{amsmath}
\usepackage{setspace}
\usepackage{epsfig}
\usepackage{cleveref}
\usepackage[font=footnotesize,labelfont=bf]{caption}
\usepackage[caption=false]{subfig}
\allowdisplaybreaks
\usepackage{tikz}

\def\del{\partial}
\newcommand{\beq}{\begin{equation}}
\newcommand{\eeq}{\end{equation}}
\newcommand{\ga}{\lower.7ex\hbox{$\;\stackrel{\textstyle>}{\sim}\;$}}
\newcommand{\la}{\lower.7ex\hbox{$\;\stackrel{\textstyle<}{\sim}\;$}}

\begin{document}

\def\jcap{\ref@jnl{J. Cosmology Astropart. Phys.}}

\begin{flushright}
{\tt UMN-TH-3903/19, FTPI-MINN-19/26, LPT-Orsay-19-37}  \\
\end{flushright}

\vspace{0.2cm}
\begin{center}
{\bf {\large Inflation and Leptogenesis in High-Scale Supersymmetry}}
\end{center}

\vspace{0.1cm}

\begin{center}{
{\bf Kunio Kaneta}$^{a}$,
{\bf Yann Mambrini}$^{b}$,
{\bf Keith~A.~Olive}$^{a}$ and
{\bf Sarunas~Verner}$^{a}$}
\end{center}

\begin{center}
{\em $^a$ William I. Fine Theoretical Physics Institute, School of
 Physics and Astronomy, \\ University of Minnesota, Minneapolis, MN 55455,
 USA}\\[0.2cm]
{\em $^b$Laboratoire de Physique Th\'eorique 
Universit\'e Paris-Sud, F-91405 Orsay, France}
 
 \end{center}

\vspace{0.1cm}
\centerline{\bf ABSTRACT}
\vspace{0.1cm}
{\small No-scale supergravity provides a successful framework
for Starobinsky-like inflation models. Two classes of models
can be distinguished depending on the identification
of the inflaton with the volume modulus, $T$ (C-models), or a matter-like
field, $\phi$ (WZ-models). When supersymmetry is broken, the inflationary potential may be perturbed, placing restrictions on the form
and scale of the supersymmetry breaking sector. 
We consider both types of inflationary models in the context of high-scale supersymmetry. We further distinguish between models in which the gravitino mass is below and above the inflationary scale.  We examine the mass spectra of the inflationary sector. We also consider in detail mechanisms for leptogenesis for each model when a right-handed neutrino sector, used in the seesaw mechanism to generate neutrino masses, is employed. 
In the case of C-models, reheating occurs via inflaton decay to 
two Higgs bosons. However, there is a direct decay channel to the
lightest right-handed neutrino which leads to non-thermal leptogenesis. In the case of WZ-models, in order to achieve reheating, we associate the matter-like inflaton with one of the right-handed sneutrinos whose decay to the lightest right handed
neutrino simultaneously reheats the Universe and generates the 
baryon asymmetry through leptogenesis. 
 } 

\vspace{0.2in}

\begin{flushleft}
November 2019
\end{flushleft}
\medskip
\noindent

\newpage

\section{Introduction}

There are many motivations for supersymmetry including the solution to the hierarchy problem  \cite{Maiani:1979cx}, gauge coupling unification \cite{Ellis:1990zq},
the stability of the Higgs vacuum \cite{Ellis:2000ig}, radiative electroweak symmetry breaking \cite{ewsb}, and viable dark matter candidates~\cite{ehnos}. 
Supersymmetry also aids in the construction of inflationary models \cite{cries} allowing naturally for flat directions suitable for inflation and keeping radiative corrections in check. Indeed, the natural framework for formulating supersymmetric models of inflation is that of supergravity \cite{nost}.
However, generic supergravity models often induce what is known as the $\eta$ problem \cite{eta}, which is easily addressed in a no-scale supergravity framework~\cite{no-scale, nsi}.

It is remarkable that the Starobinsky model based on $R + R^2$ gravity~\cite{Staro,MukhChib}, which was one of the first models of inflation, is in excellent agreement with the most recent Planck measurements~\cite{planck18} of the tensor-to-scalar ratio $r = 0.0035$ and the tilt of the scalar perturbations $n_s = 0.965$.
The Starobinsky scalar potential for a canonically-normalized inflaton field, $x$, is given by:
\begin{equation}
V = \frac{3}{4} m^2 \left(1 -e^{- \sqrt{\frac{2}{3 }} x} \right)^2 \,  ,
\label{staro}
\end{equation}
and can be easily realized in no-scale supergravity~\cite{ENO6,Avatars,eno8,eno9,ENOV1,ENOV2,ENOV3,king,KLno-scale,FKR,FeKR,adfs,EGNO4,dgmo,others,egnno1,egnno23}.
These models must contain at least two chiral fields, which we will denote as $T$, a volume modulus, and $\phi$, a matter-like field~\cite{Avatars}. The non-minimal K\"ahler potential with two chiral fields is expressed as:
\begin{equation}
K \; = \; -3 \, \ln \left(T + \overline{T} - \frac{| \phi|^2}{3} \right) \, ,
\label{K}
\end{equation}
parametrizing a non-compact $\frac{SU(2,1)}{SU(2) \times U(1)}$ coset manifold~\cite{ENO6}.
The inflationary models can be divided into two classes, in which either the volume modulus $T$ or the matter-like field $\phi$ is identified as the inflaton~\cite{Avatars,ENOV1}. Depending on the specific model, an additional chiral multiplet may be necessary to break supersymmetry \cite{EGNO4,dgmo}.

The scale of supersymmetry breaking is usually assumed to be near the weak scale. 
In that case, one easily resolves the issues that motivate supersymmetry in the first place.
However, with the exception of the hierarchy problem, the problems discussed in the beginning of this section can also be resolved in the context of high-scale supersymmetry. For example, gauge coupling unification in high-scale supersymmetry has been shown to be effective in SO(10) models of grand unification~\cite{egko}. To be more precise, it is known
that gauge coupling unification also occurs in non-supersymmetric models SO(10) models when the unified gauge symmetry is broken down to the Standard Model (SM) gauge group through an intermediate scale gauge group~\cite{GN2,moqz,mnoqz,noz,mnoz}. 
Similarly, the stability of the Higgs vacuum can be maintained in both high-scale supersymmetry~\cite{egko} and non-supersymmetric models \cite{mnoz}, when an additional scalar field  below  $10^{10}$ GeV  is present~(which can also drive
radiative electroweak symmetry breaking). 
	
Among the best studied candidates for dark matter are those arising in weak scale supersymmetric models \cite{ehnos}. In this context, R-parity conservation renders the lightest supersymmetric particle (LSP) stable.  Most phenomenological studies favor neutralino dark matter models, but models with a gravitino LSP have also been considered~\cite{pp,nos,ehnos,oss,eoss5,0404231,stef,buch,rosz,covi}. 
However, both the detection of supersymmetric particles at the LHC~\cite{nosusy} and dark matter direct-detection experiments searching for neutralino dark matter, such as LUX~\cite{LUX}, PandaX-II~\cite{PANDAX}, and XENON1T~\cite{XENON}, remain elusive. 

It is possible that the supersymmetry breaking scale is beyond
the reach of the LHC, and that the corresponding scattering cross sections for multi-TeV neutralinos are below the current detection limits, as is the case in some variants of the constrained minimal supersymmetric standard model~(CMSSM)~\cite{cmssm}.
Alternatively, supersymmetry breaking may occur at the PeV scale as in models of pure gravity mediation with a wino or higgsino dark matter candidate~\cite{pgm,eioy}.

These considerations motivate us to explore models with high-scale supersymmetry breaking~\cite{hssusy,egko}. Indeed, it is possible to construct viable models with a significantly higher supersymmetry breaking scale so that all the superpartners, except for the gravitino, lie above the inflationary scale~\cite{eev,dgmo,dgkmo,kmo}. In this case, the gravitino with a mass of order $m_{3/2} \gtrsim 0.1 \, \text{EeV}$ may play the role of dark matter. Its production occurs through the reheating process after inflation~\cite{nos,ehnos,ekn,Moroi:1995fs,enor,Giudice:1999am,bbb,egnop}. In weak-scale supersymmetry, single gravitinos together with other
supersymmetric particles can be produced from scattering processes, e.g., gluon + gluon $\to$ gravitino + gluino. However, in high-scale supersymmetry models, gravitinos must be produced in pairs~\cite{bcdm}, and this process is highly sensitive to the maximum temperature. Therefore, to obtain the correct gravitino relic density, we require a relatively high reheating temperature $T_{\rm RH} \gtrsim 10^{10} \, \text{GeV}$~\footnote{In fact, once non-instantaneous reheating is considered \cite{gmop}, it is the maximum temperature attained that gives 
the largest contribution to the dark matter abundance.}.

It is also important to note that the existence of dark matter in non-supersymmetric SO(10) models is also possible when the intermediate scale gauge group is broken via a {\bf 126} dimensional representation as a $Z_2$ discrete symmetry (similar to $R$-parity) is preserved~\cite{moqz,mnoqz,noz,mnoz,so10dm,enoz}. 

In this paper, we consider high-scale supersymmetry models in conjunction with no-scale Starobinsky-like models of inflation \cite{dgmo}. We discuss inflationary models based on a non-compact $\frac{SU(2, 1)}{SU(2) \times U(1)}$ K\"ahler potential~(\ref{K}), where either the volume modulus $T$ or a matter-like field $\phi$ is driving inflation. The two types of models are distinguished by their couplings to the Standard Model (which leads to different reheating mechanisms~\cite{EGNO4}) and the supersymmetry breaking sector. 

We extend our high-scale supersymmetry framework and incorporate various models of leptogenesis~\cite{FY,leptogenesis}.
This is accomplished by introducing a right-handed neutrino sector. The small left-handed neutrino masses are obtained via the classical seesaw mechanism~\cite{seesaw}, which leads to lepton number violation. Most importantly, the decay of the heavy right-handed neutrinos into Higgs bosons and leptons produce a lepton asymmetry, which is subsequently converted to a baryon asymmetry by sphaleron transitions~\cite{spha1, spha2}. In this paper we focus on models of non-thermal leptogenesis~\cite{FY}. In this case, 
the inflaton decays directly to a right handed neutrino which is out-of-equilibrium if its
mass is larger than the reheating temperature $T_{\rm RH}$. The subsequent out-of-equilibrium decay of the right-handed neutrino then produces the lepton asymmetry.

The structure of this paper is as follows.
We first review how the Starobinsky-like inflation models arise in no-scale supergravity. In section \ref{lepto}, we review the basics of leptogenesis
as needed in our inflationary context. 
We then consider separately the case where the inflaton is associated with the $T$ field (section \ref{c-type})
or with $\phi$ (section \ref{wz-type}). Within each case, we distinguish models in which the gravitino mass is below and above the inflationary scale. Furthermore, in each case, we discuss the mechanism for reheating,
leptogenesis, and dark matter (in sections \ref{lctype} and \ref{lwztype}). 
Our conclusions are given in section \ref{summary}.

\section{No-scale Starobinsky Models of Inflation}
The Starobinsky model of inflation can be realized in a no-scale supergravity framework by considering the K\"ahler potential form, given by Eq.~(\ref{K}), and combining it with a specific choice of a superpotential. If we consider the Cecotti superpotential form~\cite{Cecotti}:
\begin{eqnarray}
	W_{\rm C} &=& \sqrt{3} \, m \, \phi\left(T-\frac{1}{2}\right),
	\label{Cpot}
\end{eqnarray}
where the inflaton is associated with the volume modulus $T$. When the vacuum expectation value of a matter-like field is fixed to $\langle \phi \rangle = 0$ by introducing the higher-order stabilization terms in the K\"ahler potential~(\ref{K})~\cite{EKN3,Avatars}, as discussed later in this section, we obtain the Starobinsky inflationary potential~(\ref{staro}) in terms of the canonically-normalized field $x$, given by the field redefinition $T =\frac12 e^{\sqrt{2/3} x}$.

Similarly, if we consider the Wess-Zumino form for the superpotential~\cite{ENO6}:
\begin{eqnarray}
	W_{\rm WZ} &=& m \, \left(\frac{\phi^2}{2}-\frac{\phi^3}{3\sqrt3}\right),	
	\label{WZpot}
\end{eqnarray}
where the inflaton is associated with a matter-like field $\phi$, and we stabilize the volume modulus dynamically at its vacuum expectation value of $\langle T \rangle = \frac{1}{2}$, the Starobinsky inflationary potential is obtained by making the canonical field redefinition $\phi = \sqrt{3} \tanh (x/\sqrt{6})$.

In both models, the scale of inflation is characterized by a single mass scale $m$, which determines the amplitude of density fluctuations, $A_s \simeq 2 \times 10^{-9}$, as measured by Planck~\cite{planck18}.
For $N_{*} = 55$, where $N_{*}$ is the number of e-foldings before the end of inflation, the mass scale corresponds to $m \simeq 1.2 \times 10^{-5} \, M_P \simeq 3 \times 10^{13}$ GeV \cite{ENO6}, and we use this value throughout this paper.~\footnote{In this paper, we work in units of the reduced Planck mass $M_P = 1/\sqrt{8\pi G_N} \simeq 2.4 \times 10^{18} \, \text{GeV}$, unless explicitly noted.}
If we combine the K\"ahler potential~(\ref{K}) with either of the superpotential forms~(\ref{Cpot}) or~(\ref{WZpot}), we find that the parameter $m$ can be identified with the mass of the canonically-normalized inflaton field. Therefore, in some models of leptogenesis it seems natural to identify the inflaton with one of the right-handed sneutrinos~\cite{eno8}, which then decays into right-handed neutrinos responsible for leptogenesis.

In fact, the two models listed above are simply two examples of a wide class of superpotential models which all generate the same scalar potential~\cite{Avatars,ENOV1}. In
the absence of supersymmetry breaking, one can show that these classes can be related by the underlying 
non-compact $\frac{SU(2,1)}{SU(2) \times U(1)}$ no-scale symmetry \cite{ENOV1}. Once the theory is coupled 
to matter and supersymmetry is broken, this symmetry is broken and different models will have different phenomenologies \cite{EGNO4}.

Neither of the superpotentials~(\ref{Cpot}) nor~(\ref{WZpot}) are responsible for supersymmetry breaking. In the absence of supersymmetry breaking, the minimum of the scalar potential is located at $\langle T \rangle = \frac{1}{2}$ and $\langle \phi \rangle =0$. Therefore, we need to extend our models and incorporate supersymmetry breaking. In the Cecotti model~(\ref{Cpot}), supersymmetry can be broken by introducing a Polonyi field $z$~\cite{pol} with superpotential:
\begin{eqnarray}
\label{ppot}
	W_{\rm P} &=& {\tilde m}(z+b) \, ,
\end{eqnarray}
where $b$ is a constant. 
It is important to note that the presence of a Polonyi field will shift the minimum, and in Section
\ref{c-type} we discuss this in more detail. If we consider the combined superpotential $W = W_{C} + W_{P}$, we obtain an upper limit ${\tilde m} < m/2$ for viable solutions with a Minkowski vacuum. The gravitino mass in this case is given by~$m_{3/2}={\tilde m}/\sqrt{3}$, and is lighter than the
inflaton and can be a good dark matter candidate.

In the Wess-Zumino model~(\ref{WZpot}), the superpotential is a function of a matter-like field $\phi$ only.
In this case, one does not need to introduce an additional Polonyi field, and supersymmetry breaking occurs by introducing a constant~$\lambda_1$ in the superpotential. More generally, we can add the following superpotential term~\cite{ENOV2,ENOV3}:
\beq
W_{\rm SSB} = \lambda_1 - \lambda_2 \, \left(2T - \frac{\phi^2}{3} \right)^3,
\eeq
which generates supersymmetry breaking through an $F$-term, which is given by $F_{T} = \lambda_1 + \lambda_2$. In this case, the gravitino mass is given by $m_{3/2} = \lambda_1 - \lambda_2$, and the vacuum energy density is expressed as $V_0 = 12 \, \lambda_1 \lambda_2$~\cite{enno}, which vanishes if either $\lambda_1$ or $\lambda_2$ is set to zero. For simplicity, we consider models with $\lambda_2 = 0$, and supersymmetry breaking is achieved by a constant $\lambda_1$, whose relative size is not restricted by the inflaton mass, $m$. We note that it is possible to add a linear term in the Wess-Zumino superpotential~(\ref{WZpot}), which behaves as a Polonyi-like field~\cite{king}. However, this model has a strict upper bound on the gravitino mass of $m_{3/2} \lesssim 10^6 \, \text{GeV}$, and it is not valid for high-scale supersymmetry models. Another possibility is to introduce a Polonyi sector to the Wess-Zumino models~\cite{EGNO4}, however in that case the inflationary potential is affected and the possibility for inflation becomes limited.

In both models, the phenomenological aspects for the limits $m_{3/2}>m$ and $m_{3/2}<m$ are distinct, therefore, we consider the four cases separately. The classification is shown in Table~\ref{tab:models}.

\begin{table}[!ht]
\centering
	\begin{tabular}{c|cc}
		& $W_{\rm C}$ & $W_{\rm WZ}$ \\\hline
		$m_{3/2}<m$ & C-1 & WZ-1 \\
		$m_{3/2}>m$ & C-2 & WZ-2
	\end{tabular}
	\caption{The classification of the high-scale supersymmetry models considered here.}
	\label{tab:models}
\end{table}

It is crucial to note that for both types of models, stabilization of some fields is necessary and can be achieved dynamically by introducing higher-order correction terms in the K\"ahler potential. We consider the following general K\"ahler potential form:
\begin{eqnarray}
\label{K2}
	K &=& -3\ln\left[T+\overline T+f(T,\overline T) - \frac{|\phi|^2 + |z|^2}{3} +g(\phi,\overline\phi)+h(z,\overline z)\right],
\end{eqnarray}
where
\begin{align}
& \qquad f(T,\overline T)=0,          &  g(\phi,\overline\phi)  &=\frac{|\phi|^4}{\Lambda_\phi^2},              &  h(z,\overline z)&=\frac{|z|^4}{\Lambda_z^2}, \\
\shortintertext{\vspace*{.3cm}for the Cecotti models, and}
f(T,\overline T)&=\frac{(T+\overline T-1)^4}{\Lambda_{T}^2}+\frac{d(T-\overline T)^4}{\Lambda_{T}^2},         &  g(\phi,\overline\phi)  &=0,             &  h(z,\overline z)&=0,
\label{d}
\end{align}
for the Wess-Zumino models. A more detailed discussion related to stabilization can be found in~\cite{Avatars, dine,Dudas:2006gr,klor,dlmmo,eioy,nataya,ADinf,ego,EGNO4}.  For all correction terms, $\Lambda_{T}$, $\Lambda_\phi$, and $\Lambda_z$ are associated with the corresponding field stabilization, and they are assumed to be below the Planck scale $M_P$.
As we discuss in the next section, due to supersymmetry breaking in Cecotti-type models, the VEV of a matter-like field $\phi$ is no longer zero, and to avoid the uplifting of Minkowski vacuum by strong stabilization effects, we impose the constraint $\langle \phi \rangle \lesssim \Lambda_{\phi}$.

One of the features of all the models discussed here
is a high supersymmetry breaking scale. As noted earlier,
we require that all sparticle masses are larger than 
the inflationary scale given by the inflaton mass 
with the possible exception of the 
gravitino. More specifically, we must (at least in some 
cases) generate a hierarchy between gaugino masses and the
gravitino mass. 
Gaugino masses
are given by:
\beq
m_{1/2} = \left| \frac{1}{2} e^{G/2}\frac{\bar{f}_{z}}{{\rm Re} f} (G^{-1})^z_z G^z \right| \simeq  \left| \frac{1}{2} m_{3/2} \frac{\bar{f}_{z}}{{\rm Re}\,f}  \right| \,,
\label{gaugino}
\eeq
where $G = K + \log |W|^2$ is the K\"ahler function, 
$f_{\alpha\beta} = f \, \delta_{\alpha\beta} $ is the gauge kinetic function, $f_{\alpha \beta} F^\alpha_{\mu\nu} {F^\beta}^{\mu\nu}$. In the case of a strongly stabilized Polonyi field, we can write $ f = f_0 + f_1 \, z/\Lambda_z$
where $f_0 \sim 1/g^2$ is related to the gauge coupling, 
and the VEV of $z$ is proportional to $\Lambda_z^2/M_P \ll M_P$ (see below).
Then ${\bar{f}_z}/f = f_z/f \sim g_0^2 f_1/\Lambda_z$, and the gaugino mass is $m_{1/2} \sim g_0^2 f_1 m_{3/2} M_P/\Lambda_z \gg m_{3/2}$. Scalar masses may then receive contributions from
gaugino loops so that $m_0^2 \, \propto \, m_{1/2}^2/16\pi^2$ \cite{dgmo}.
\section{Models of Leptogenesis}
\label{lepto}

Before we discuss leptogenesis in the context of the 
two inflationary paradigms, we first review some of the general formalism for generating a baryon asymmetry from a lepton asymmetry induced by the out-of-equilibrium decay of a heavy right-handed neutrino. For the most part, we concentrate on non-thermal leptogenesis~\cite{FY}. We begin with the introduction of 
right-handed neutrinos and their role in the seesaw mechanism for generating neutrino masses. Later, we will associate one of the right-handed neutrinos with a fermionic partner of the inflaton~\cite{eno8}. We then give the basic formulae for generating a lepton asymmetry from the decays of right-handed neutrinos, and its 
subsequent conversion to a baryon asymmetry through sphaleron interactions. 

\subsection{The Seesaw Mechanism}
We begin our discussion by recalling the general features of the seesaw mechanism \cite{seesaw,lept2}. We introduce three generations of heavy right-handed neutrinos, that will produce a lepton asymmetry and generate the masses of the light neutrinos via the seesaw mechanism. In this case, the new terms in the Lagrangian are given by:
\begin{equation}
\label{lagsee}
\mathcal{L} \supset -y_{i \alpha} {\bar N}_i L_{\alpha}  H_u - \frac{1}{2} {\bar N^c_i} M_i N_i + \text{h.c.} \, ,
\end{equation}
where $\alpha = e, \mu, \tau$, $i = 1, 2, 3$, and the Yukawa couplings are given by a $3 \times 3$ matrix $y$. For simplicity, we assume that the right-handed neutrino mass matrix $M$ is diagonal.~\footnote{Even if we do not assume that the right-handed neutrino matrix $M$ is diagonal, we can always diagonalize it by introducing a unitary matrix $U_R$.} To obtain the Dirac mass matrix via the seesaw mechanism, we use the following expression~\cite{seesaw}:
\begin{equation}
\label{dir1}
M_{\nu} = m_D^{T} \, M^{-1} \, m_D,
\end{equation}
where $M^{-1}$ is the inverse of the diagonal right-handed neutrino mass matrix, and $m_D = y \langle H_u \rangle$, where $\langle H_u \rangle = v \sin{\beta}$ with $v = 174 \, \text{GeV}$. The left-handed neutrino masses are obtained by diagonalizing the mass matrix $M_{\nu}$~(\ref{dir1}) with the Pontecorvo-Maki-Nakagawa-Sakata (PMNS) matrix $U$:
\begin{equation}
\text{diag} \{m_{\nu_1}, m_{\nu_2}, m_{\nu_3} \} = U^{T} \, M_{\nu} \, U.
\end{equation}
If we integrate out the heavy right-handed neutrinos, the left-handed neutrino masses become:
\begin{equation}
\label{seesaw1}
(M_{\nu})_{\alpha \beta } = \sum_i y_{i \alpha} y_{i \beta} \frac{\langle H_u \rangle^2}{M_i}.
\end{equation}
For the consideration of non-thermal leptogenesis, we assume the 
following mass hierarchy: $2M_1 \lesssim m \ll M_2, M_3$, and lepton asymmetry will be primarily generated by the decays of the lightest right-handed neutrino $N_1$. 

If we assume that the dominant contribution to the Yukawa matrix $y$ comes from the entry $y_{3 \tau} \equiv y_3$, the seesaw formula~(\ref{seesaw1}) leads to the following mass eigenvalue:
\begin{equation}
\label{neut3}
m_{\nu_3} \simeq \frac{|y_{3}|^2 \,\langle H_u \rangle^2}{M_3},
\end{equation}
 which corresponds to the heaviest left-handed neutrino in the normal hierarchy.
 
Analogously, we can consider the case when largest entry in the Yukawa matrix $y$ is $y_{2 \mu} \equiv y_2$, which leads to:
\begin{equation}
\label{neut2}
m_{\nu_2} \simeq \frac{|y_{2}|^2 \,\langle H_u \rangle^2}{M_2}.
\end{equation}
For a normal hierarchy~(\textit{NH}) of the left-handed neutrinos, their masses are expressed as~\cite{pdg}:
\begin{equation}
\label{normhier}
m_{\nu_2} \simeq  0.0086 \,\text{eV}, \qquad m_{\nu 3} \simeq  0.0506 \, \text{eV},
\end{equation}
and $m_{\nu_1}$ is very light. For inverted hierarchy~(\textit{IH}), the neutrino masses are given by:
\begin{equation}
\label{invhier}
    m_{\nu_1} \simeq 0.0497 \,\text{eV}, \qquad m_{\nu_2} \simeq  0.0504 \, \text{eV},
\end{equation}
where $m_{\nu_3}$ is very light.

\subsection{Lepton Asymmetry from Heavy Majorana Neutrino Decays}
When the heavy right-handed Majorana neutrinos decay into leptons and Higgs bosons or their antiparticles, lepton number is violated. The lepton asymmetry $\epsilon$ is generated by the  interference between one-loop and tree diagrams of the following out-of-equilibrium decays of the lightest right-handed neutrino $N_1$:
\begin{equation}
\begin{split}
& N_{1} \rightarrow L_{\alpha} + H_{u} \\
& N_{1} \rightarrow \bar{L}_{\alpha} + \overline{H}_{u}.
\end{split}
\end{equation}
For our models of non-thermal leptogenesis we assume $2M_1 \lesssim m \ll M_{2, \, 3}$, where $m \simeq 3 \times 10^{13} \, \text{GeV}$ is the mass of the inflaton. The expression for the $CP$ asymmetry is given by~\cite{luty,CPviol}:
\begin{equation}
\label{CP1}
\epsilon \equiv \frac{\Gamma_{N_1 \rightarrow L_{\alpha} H_{u}} - \Gamma_{N_1 \rightarrow \bar{L}_{\alpha} \overline{H}_{u}}}{\Gamma_{N_1 \rightarrow L_{\alpha} H_{u}} + \Gamma_{N_1 \rightarrow \bar{L}_{\alpha} \overline{H}_{u}}} \simeq \frac{1}{8 \pi} \frac{1}{(y y^{\dagger})_{11}} \sum_{j = 2, \, 3} \text{Im} \left(y y^{\dagger} \right)_{1j}^2 \cdot f \left(\frac{M_j^2}{M_1^2} \right),
\end{equation}
where
\begin{equation}
f(x) = \sqrt{x} \left[\frac{1}{1-x}+1 - (1+x)\ln \left(\frac{1+x}{x} \right) \right].
\end{equation}
For $x \gg 1$, we use the approximation $f(x) \simeq -3/2 \sqrt{x}$, and the $CP$ asymmetry parameter~(\ref{CP1}) becomes:
\begin{equation}
\label{CP2}
\epsilon \simeq -\frac{3}{16 \pi} \frac{1}{(y y^{\dagger})_{11}} \left[ \text{Im} \left(y y^{\dagger} \right)_{12}^2 \frac{M_1}{M_2}  + \text{Im} \left(y y^{\dagger} \right)_{13}^2 \frac{M_1}{M_3}\right].
\end{equation}
If we consider the case when $y_{3 \tau} = y_3$ is the dominant contribution in the Yukawa matrix $y$, we can express the $CP$ asymmetry parameter~(\ref{CP2}) as:
\begin{equation}
\epsilon \simeq -\frac{3 \, \delta_{\text{eff}} \, |y_{3}^2|}{16 \pi} \, \frac{M_1}{M_3},
\end{equation}
where $\delta_{\text{eff}}$ is the effective $CP$-violating phase. 

Similarly, we can assume that $y_{2 \mu} = y_2$ is the largest entry in the Yukawa matrix $y$, and then $CP$ asymmetry parameter~(\ref{CP2}) becomes:
\begin{equation}
\epsilon \simeq -\frac{3 \, \delta_{\text{eff}} \, |y_{2}^2|}{16 \pi} \, \frac{M_1}{M_2}.
\end{equation}
If we then use the seesaw expression~(\ref{neut3}) or~(\ref{neut2}), we find:
\begin{equation}
\label{cpneut}
\epsilon \simeq -\frac{3 \, \delta_{\text{eff}}}{16 \pi} \cdot \frac{m_{\nu_i} \, M_1}{v^2 \, \sin^2 \beta}, 
\end{equation}
where $i = 2, 3$ for normal hierarchy, and analogously, we can find the $CP$ asymmetry parameter~(\ref{cpneut}) for inverted hierarchy, with ~$i = 1, 2$.

Eq.~(\ref{cpneut}) can be used to calculate the lepton asymmetry generated by the out-of-equilibrium decays of lightest right-handed neutrino $N_1$, and similar models were discussed in~\cite{cdo,giudicereh, leptinf1, leptinf2,egnno5}. In the next section we discuss how a lepton asymmetry is converted to a baryon asymmetry by sphaleron transitions.

\subsection{Production of Baryon Asymmetry}
We briefly discuss the mechanism which converts a lepton asymmetry into a baryon asymmetry via electroweak sphaleron interactions~\cite{spha1}. At high temperatures, the combination of baryon and lepton number $B + L$ is violated, while the anomaly-free parameter $B - L$ remains conserved. Sphaleron interactions are in equilibrium in the temperature range between~$100 \, \text{GeV}$ and~$10^{12} \, \text{GeV}$, and they convert a fraction of a non-zero $B-L$ asymmetry into a baryon asymmetry~\cite{spha2}:
\begin{equation}
\label{btol}
Y_B \simeq a \, Y_{B-L},
\end{equation}
where $Y_B = n_B/s$, $Y_{B - L} = n_{B - L}/s$, and
\begin{equation}
\label{coef}
a = \frac{8N_F + 4N_H}{22 N_F + 13 N_H},
\end{equation}
where $N_F$ is the number of fermion generations and $N_H$ is the number of Higgs doublets. In our case, we have $N_F = 3$, $N_H = 1$, and $a = 28/79$. In leptogenesis, where purely a lepton asymmetry is generated, $B-L = -L$.

For models of non-thermal leptogenesis, we impose the constraint $M_1 > T_{\rm RH}$~\footnote{To preserve the lepton asymmetry, we require that the lepton number violating interaction, which is characterized by an operator $y^2LLH_u H_u/M$, remains out-of-equilibrium when sphaleron transitions are in thermal equilibrium. It was shown in~\cite{dgkmo2}, that for high-scale supersymmetry models we must satisfy the constraint $M/|y|^2 > 10^{13.5}$, or $m_{\nu} < 0.5 \, \rm{eV}$.} (the right-handed neutrino $N_1$ must be heavier than the reheating temperature $T_{\rm RH}$), and lepton asymmetry is produced through the out-of-equilibrium decay of the lightest right-handed neutrino $N_1$. In this case, we acquire the following expression for lepton asymmetry:
\begin{equation}
Y_L \equiv \epsilon \frac{n_{N_1}}{s},
\end{equation}
and if we relate it to the baryon asymmetry number using Eq.~(\ref{btol}), we find:
\begin{equation}
\label{asy}
Y_{B} \simeq -a \, \epsilon \frac{n_{N_1}}{s}.
\end{equation}
If we combine the expressions~(\ref{cpneut}) and~(\ref{coef}) with~(\ref{asy}), and assume that for high-scale supersymmetry models we have $\tan{\beta} \simeq 1$, we obtain the following expression for the baryon asymmetry:
\begin{align}
\label{baryonasy2}
Y_{B} \simeq 7 \times 10^{-5} \, \delta_{\text{eff}} \, \frac{n_{N_1}}{s} \left(\frac{m_{\nu_i}}{0.05 \, \text{eV}} \right) \, \left(\frac{M_1}{10^{12} \, \text{GeV}} \right) , \qquad~&\text{where}~i=2, \, 3.
\end{align}

Finally, the baryon asymmetry of the Universe is given by the most recent Planck data constraints~\cite{planck18}:
\begin{equation}
\label{baryonasy3}
\eta_B = \frac{n_B - n_{\bar{B}}}{n_{\gamma}} \simeq 6.12 \times 10^{-10}, \qquad Y_B \simeq 8.7 \times 10^{-11}.
\end{equation}
\section{Cecotti-type Models of Inflation}
\label{c-type}
We begin by considering the inflationary models where the inflaton is associated with the volume modulus $T$. In particular, we consider the superpotential form $W_C$, given by Eq. (\ref{Cpot}).
However, for Cecotti-type models, one cannot introduce a constant term in the superpotential, because it shifts the original minimum to a new supersymmetry preserving AdS vacuum~\cite{EGNO4,ENOV3}. Therefore, we introduce a Polonyi sector and consider the K\"ahler potential form~(\ref{K2}) with the superpotential $W_C + W_P$, where $W_P$ is given by Eq.~(\ref{ppot}). 

The addition of a Polonyi sector shifts the scalar potential minimum to a new vacuum with broken supersymmetry. In the absence of superpotential $W_C$, the strongly stabilized Polonyi potential has a minimum at $\langle z \rangle \simeq \Lambda_z^2/2\sqrt{3}$, where we have omitted the higher-order terms in $\Lambda_z$. If we choose a constant $b \simeq 1/\sqrt{3}$, we obtain a vanishing vacuum energy density~$V = 0$.

When we consider the superpotential combination $W_C + W_P$, the VEVs of the fields $T$, $\phi$, and $z$ shift. However, the VEVs of the shifted fields will depend on whether the parameter $\Delta \equiv \tilde{m}/m >1/2$ or $\Delta < 1/2$. For the latter case, 
the supersymmetry breaking Minkowski minimum~$V = 0$ is located at:
\begin{subequations}
    \begin{align}
       \langle T\rangle &= \frac{1}{6}(4-\sqrt{1-4\Delta^2}),\label{cshiftst}\\
	\langle \phi\rangle &= \pm\left(\frac{1}{2}-\frac{1}{2}\sqrt{1-4\Delta^2}\right)^{1/2},
	\label{cshiftsf} \\
	\langle z\rangle &= \mp\frac{\Lambda_z^2}{6\sqrt{6}\Delta}(1-\sqrt{1-4\Delta^2})^{1/2},
	\label{cshiftsz}
\end{align}
\vspace{-1.0cm}
	\begin{gather}
	b = \pm\frac{\sqrt{6}}{18\Delta}(2+\sqrt{1-4\Delta^2})(1- \sqrt{1-4\Delta^2})^{1/2} \,,
	\label{cshiftsb}
\end{gather}
\end{subequations}
where the the higher-order terms in $\Lambda_z$ have been omitted. For small values of $\Delta$, the VEVs~(\ref{cshiftst} - \ref{cshiftsb}) can be expanded to:
\begin{equation}
\label{vevseq}
\langle T \rangle \simeq \frac{1}{2} + \frac{\Delta^2}{3}, \qquad 
\langle \phi \rangle \simeq \pm \Delta, \qquad
\langle z \rangle \simeq \pm \frac{\Lambda_z^2}{6 \sqrt{3}}, \qquad
b \simeq \pm \frac{1}{\sqrt{3}} \mp \frac{\Delta^2}{6 \sqrt{3}},
\end{equation}
which agrees with the previous results~\cite{EGNO4,dgmo}. One can see from Eqs.~(\ref{cshiftst} - \ref{cshiftsb}) that the largest possible value is $\Delta = 1/2$, and larger values of the parameter $\Delta$ lead to a positive vacuum energy density of order $\tilde{m}^2 \Lambda_z^2$. However, if we modify the Cecotti superpotential~(\ref{Cpot}), it is possible to accommodate the values $\Delta > 1/2$, and we discuss this possibility in Sec.~\ref{sec:C-2}.

Using the vacuum expectation values~(\ref{cshiftst} - \ref{cshiftsb}), we consider two separate Cecotti-type models: C-1 models, where the gravitino is lighter than the inflaton, $m_{3/2} < m$, and C-2 models, where the gravitino is heavier than the inflaton, $m_{3/2} > m$.
\subsection{C-1 Models}
\label{sec:C-1}
For C-1 models, the gravitino plays the role of the dark matter candidate, and all other sparticles are taken to be heavier than the inflaton field~\cite{dgmo}. It was shown in the previous section that  C-1 models are valid when $\Delta = \sqrt{3} m_{3/2}/m <1/2$, or $m_{3/2} < m/2 \sqrt{3}$, which shows that the gravitino is lighter than the inflaton.

We begin by considering the relevant supergravity Lagrangian for the scalar fields:
\begin{eqnarray}
\label{lagsca}
	{\cal L} &=& -G_{i \bar{j}} \,\del_\mu\phi^i\del^\mu\phi^{\bar{j}} - V,
\end{eqnarray}
where $V$ is the effective scalar potential generated by $F$-term contributions, and we defined the K\"ahler function as $G = K + \ln{W} + \ln{\overline{W}}$. From the K\"ahler potential form~(\ref{K2}) and the Lagrangian~(\ref{lagsca}), we find that the canonically-normalized fields are expressed as:
\begin{eqnarray}
\label{canT}
	T &=& \frac{1}{2}\left(e^{\sqrt{\frac{2}{3}}T_R}+i \, \sqrt{\frac{2}{3}}T_I\right),\\
	\phi &=&\frac{1}{\sqrt2}(\phi_R+i\phi_I),\\
	z &=& \frac{1}{\sqrt2}(z_R+iz_I),
\end{eqnarray}
where $T_R$ is the inflaton.
Then, if we assume that $\Delta \ll 1$ and neglect the higher-order contributions, we obtain the following masses for the canonically normalized fields:
\begin{subequations}
    \begin{align}
     	&m_{S_{1R}}^2 \simeq \left(1+\frac{\Delta}{\sqrt{3}}\right)m^2, 
     	& m_{S_{1R}}& \simeq \left(1+\frac{\Delta}{2\sqrt{3}}\right)m,\\
     	&m_{S_{2R}}^2 \simeq \left(1-\frac{\Delta}{\sqrt{3}}\right)m^2,
     &	m_{S_{2R}}& \simeq \left(1-\frac{\Delta}{2\sqrt{3}}\right)m, \\
	&m_{S_{1I}}^2 \simeq \left(1-\frac{\Delta}{\sqrt{3}}\right)m^2,
   &m_{S_{1I}}& \simeq \left(1 - \frac{\Delta}{2\sqrt{3}}\right)m,\\
	&m_{S_{2I}}^2 \simeq \left(1 + \frac{\Delta}{\sqrt{3}}\right)m^2,  &m_{S_{2I}}& \simeq \left(1 + \frac{\Delta}{2\sqrt{3}}\right)m, \\
	&m_{z_R}^2 \simeq 12 m^2 \frac{\Delta^2 }{\Lambda_z^2}, 
	&m_{z_R}& \simeq 2 \sqrt{3} m \frac{\Delta}{\Lambda_z},\\
	& m_{z_I}^2 \simeq 12 m^2\frac{\Delta^2}{\Lambda_z^2},
	& m_{z_I} &\simeq 2 \sqrt{3} m \frac{\Delta}{\Lambda_z}, \\
	& m_{3/2} \simeq m \frac{\Delta}{\sqrt{3}}   \, ,
    \end{align}
\end{subequations}
where the eigenstates $S_{1,2R}$ correspond to equal mixtures of the real states $T_R$ and $\phi_R$, and the eigenstates $S_{1,2I}$ correspond to equal mixtures of the imaginary states $T_I$ and $\phi_I$, given by:
\begin{align}
S_{1R} &\simeq \frac{1}{\sqrt{2}} \left(T_R - \phi_R \right), & S_{2R} &\simeq  \frac{1}{\sqrt{2}} \left(T_R + \phi_R \right), \\
S_{1I} &\simeq \frac{1}{\sqrt{2}} \left(T_I - \phi_I \right), & S_{2I} &\simeq  \frac{1}{\sqrt{2}} \left(T_I + \phi_I \right).
\end{align}

It is important to note that the Polonyi field mixing with fields $T_{R,I}$ and $\phi_{R,I}$ was neglected. In order to ensure that the entropy production from the Polonyi sector is sufficiently small~\cite{ego}, and to avoid the production of the particles $z_{R, I}$ from the inflaton decays, we assume that $2 \sqrt{3} \Delta \lesssim\Lambda_z \lesssim 10^{-2}$.

Next, we consider the Lagrangian terms for the left-handed chiral fermions $\chi_L$,  given by:
\begin{eqnarray}
	{\cal L} & \supset & - g_{i \bar{j} }\overline\chi^i_L \slashed{D} \chi^{\bar{j}}_L - \left( \frac{1}{2} m_{ij} \overline{\chi}^{i}_L\chi_L^j+\text{h.c.} \right),
\end{eqnarray}
with:
\begin{eqnarray}
	g_{i \bar{j}} &=& G_{i\bar{j}} - \frac{1}{3}G_iG_{\bar{j}},\\
	m_{ij} &=& G_{ij}+\frac{1}{3}G_iG_j-\Gamma^k_{ij}G_k,
\end{eqnarray}
where we subtracted the Goldstino mode. Here, we defined, $G_i = \partial G/\partial \phi^i$ and $G_{ij} = \partial^2 G/\partial \phi^i \partial \phi^j$, where $\phi^i$ is the scalar partner of $\chi_L^i$, and $\Gamma^k_{ij}$ are the Christoffel symbols~(for a more detailed discussion, see~\cite{Ferrara:2016ntj}). It should be noted that in general $g_{i \bar{j}}$ is not the identity matrix, and relevant fields should be canonically normalized. In our case, the Goldstino is identified with the fermionic partner of the Polonyi field $z$, and the physical masses of the remaining fermions are given by:
\begin{eqnarray}
	m_{\chi_{1}} &\simeq& \left(1+\frac{\Delta \Lambda_z^2}{18\sqrt{3}}\right)m, \\	
	m_{\chi_{2}} &\simeq& \left(1-\frac{\Delta \Lambda_z^2}{18\sqrt{3}}\right)m.
\end{eqnarray}
As in the scalar field case, the fermion mass eigenstates are a mixture of eigenstates $\chi_T$ and $\chi_{\phi}$, which are related to $\chi_1$ and $\chi_2$ by:
\begin{eqnarray}
    \chi_1 \simeq \frac{1}{\sqrt2}(\chi_T+\chi_\phi),&&
    \chi_2 \simeq \frac{1}{\sqrt2}(\chi_T-\chi_\phi).
\end{eqnarray}

\begin{figure}[!ht]
\centering
\includegraphics[scale=.55]{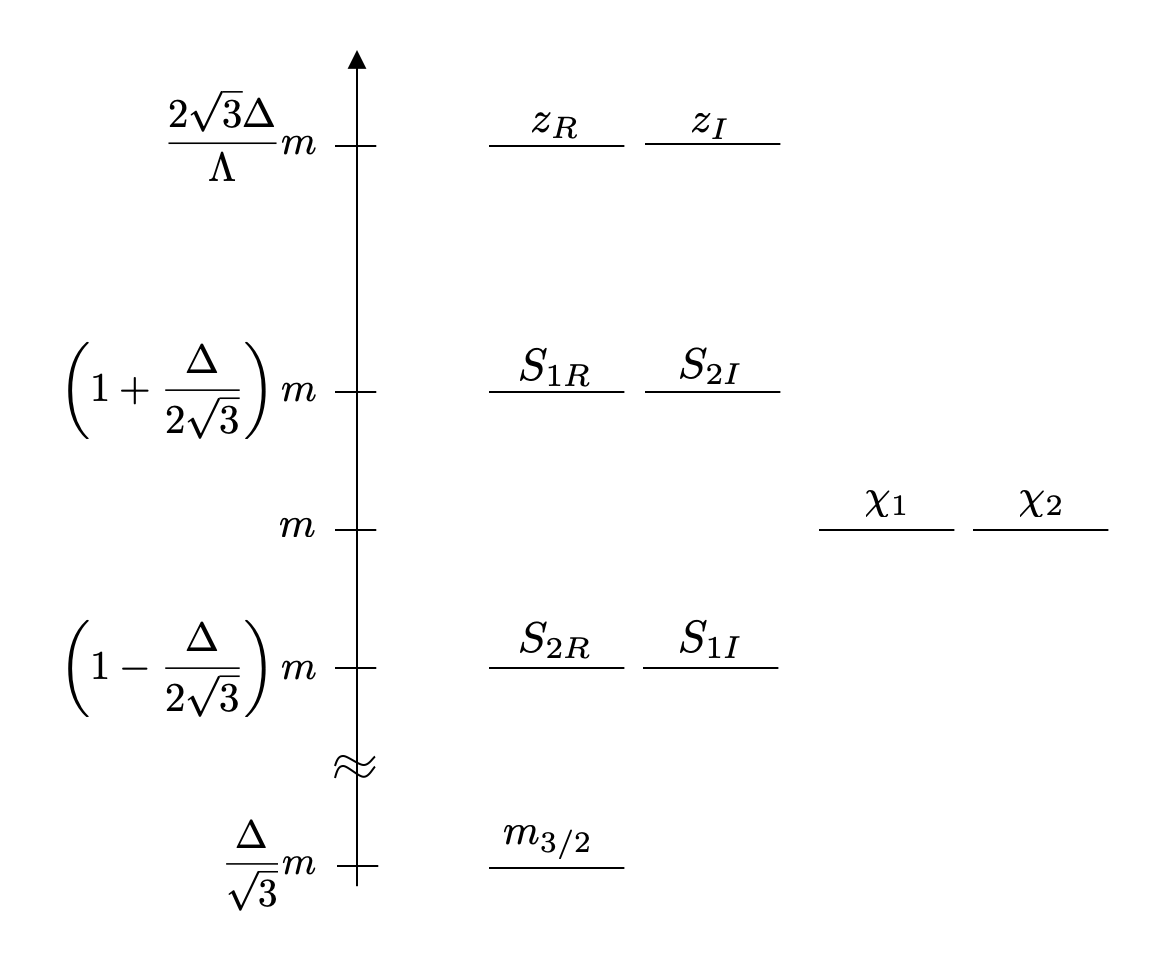}
\caption{Mass spectrum of model C-1.}
\label{spectrumC1}
\end{figure}

Although the states $S_{1R}$ and $S_{2I}$ are heavier than the scalar states $S_{2R}$ and $S_{1I}$ or the fermion states $\chi_{1, \, 2}$, the decays to lighter states are kinematically forbidden. For example, if we would consider the decay channel of the state $S_{1R}$ into a fermion $\chi_{1, \, 2}$ and a gravitino, the mass splitting of the states is smaller than the gravitino mass, i.e.,  $m_{S_{1R}} - m_{\chi_{1, 2}} \simeq m_{3/2}/2$, and the decay is kinematically forbidden~\cite{nos,nop}. Finally, we show the mass spectrum for model C-1 in Fig.~\ref{spectrumC1}.
\subsection{C-2 Models}
\label{sec:C-2}
For C-2 models, the inflaton decay does not produce gravitinos because the gravitino mass is heavier than the inflaton, $m_{3/2} > m$, and we need to consider a different dark matter candidate. In this paper we mostly focus on inflation and leptogenesis, and studies related to different dark matter candidates are left for future work.

If we look at the VEVs of fields $T$, $\phi$, and $z$, given by Eqs. (\ref{cshiftst} - \ref{cshiftsb}), we see that viable  models with Minkowski vacua impose the constraint $\Delta < 1/2$.
However, this constraint can be avoided if we modify the superpotential~(\ref{Cpot}). For example, consider the superpotential:
\begin{equation}
\label{cec1}
W_{C} = \sqrt{3} m \, \phi \left(T - \frac{1}{2} + g \, \phi^2 \right),
\end{equation}
where we introduced the term $g \, \phi^2$.
Because we dynamically stabilize a matter-like field to $\phi = 0$ during inflation, the additional term in (\ref{cec1}) does not affect the inflationary potential. The introduction of a new term alters the solutions for the field VEVs, which become:
\begin{subequations}
\begin{align}
\label{vev1}
\langle T \rangle &= \frac{4 - 63 g + (9 g-1) \sqrt{1 - 4 \Delta ^2 (1 - 18g)}}{6-108 g}, \\
\label{vev2}
\langle \phi \rangle  &= \pm \sqrt{\frac{1-\sqrt{1 - 4 \Delta ^2 \left(1 - 18 g\right)}}{2(1- 18 g)}}, \\
\label{vev3}
\langle z \rangle &= \pm \frac{\Lambda ^2 \sqrt{1-\sqrt{1-4 \Delta ^2 (1-18 g)}}}{6 \sqrt{6} \Delta\sqrt{1-18 g}},
\end{align}
\vspace{-1.0cm}
\begin{gather}
\label{vev4}
b = \pm \frac{\sqrt{1-\sqrt{1-4 \Delta ^2 (1-18 g)}} \left(2 + \sqrt{1-4 \Delta ^2 (1-18 g)}\right)}{3 \sqrt{6} \Delta  \sqrt{1- 18g}}.
\end{gather}
\end{subequations}
In the limit, $g\to 0$, we recover the solutions~(\ref{cshiftst} - \ref{cshiftsb}), given for model C-1. However, when $\Delta > 1/2$, we see from Eqs.~(\ref{vev1} - \ref{vev4}), that in order to maintain the real values for the shifted VEVs, we must satisfy the following constraints on $g$:
\begin{equation}
1 - 18g \geq 0, \qquad \sqrt{1 - 4 \Delta^2(1 - 18g)} \geq 0.
\end{equation}
We find the following inequality for the constant $g$:
\begin{equation}
g \geq \frac{1}{18} - \frac{1}{72 \Delta^2},
\end{equation}
which for large values of $\Delta$ can be approximated to $g \simeq 1/18$. In most cases to obtain a viable solution with a Minkowski vacuum at the minimum, we will need to choose a value of $g$, which is very close to the upper bound $g \simeq 1/18$, otherwise the potential is uplifted resulting in a positive vacuum energy density.

\begin{figure}[!ht]
\centering
\includegraphics[scale=.7]{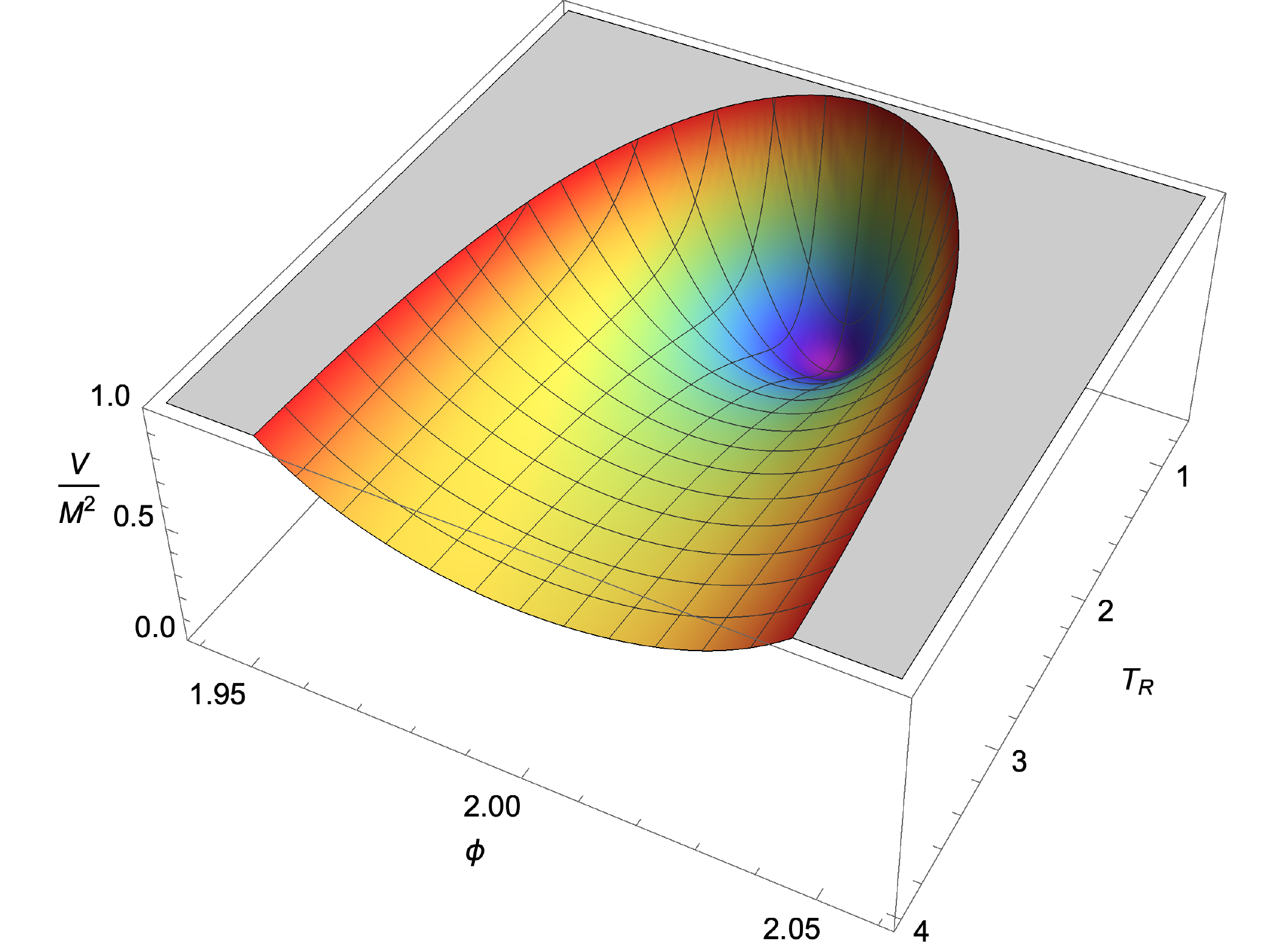}
\caption{Realization of the Starobinsky-like inflationary potential in model C-2 for $\Delta = 2$. The minimum of the potential is located at $\langle T_R \rangle \simeq 1.07$ and $\langle \phi \rangle \simeq 2$. }
\label{plot1}
\end{figure}

\begin{figure}[!ht]
\centering
\includegraphics[scale=.6]{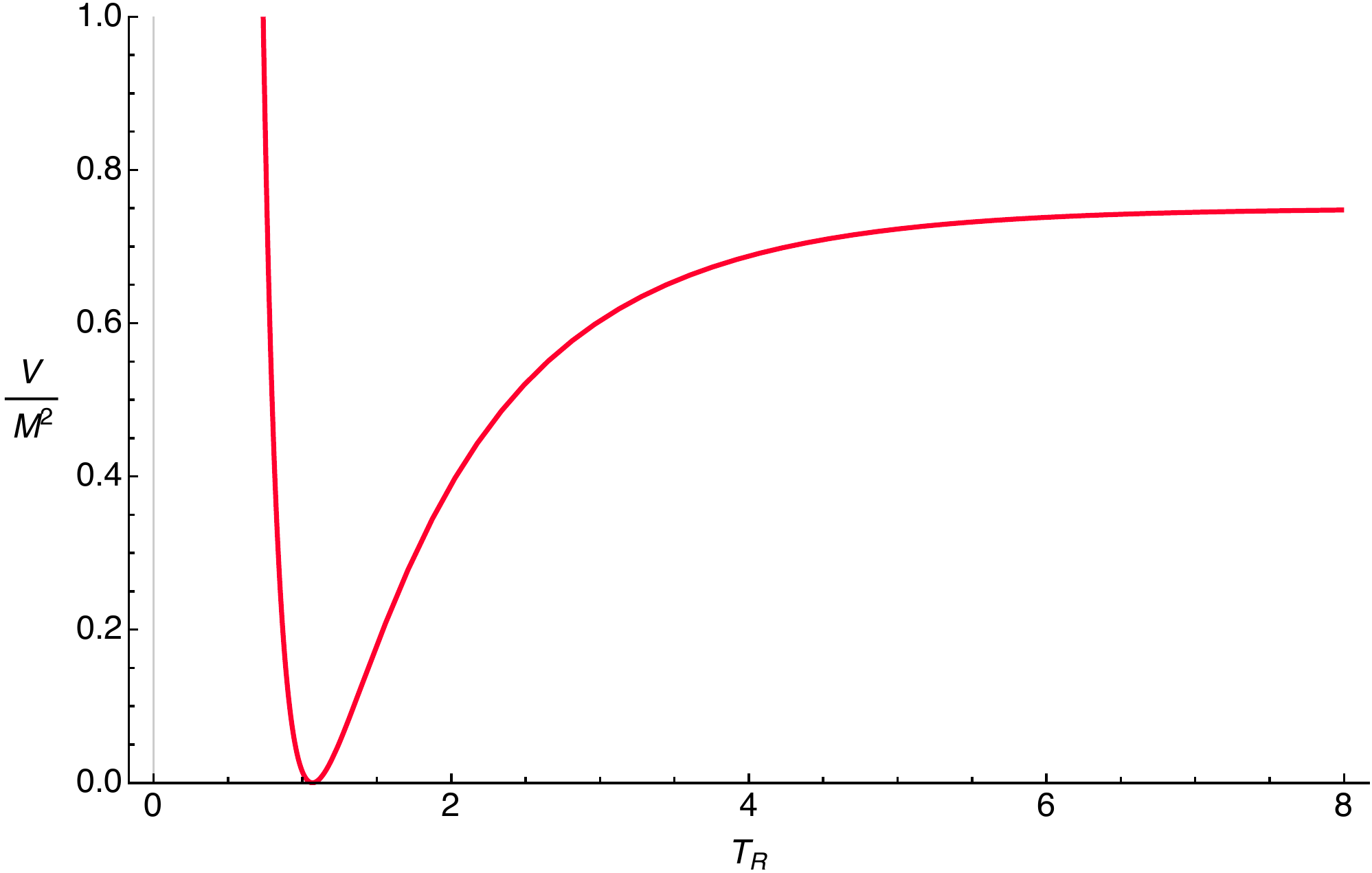}
\caption{Realization of the Starobinsky-like inflationary potential in model C-2 for $\Delta = 2$ when a matter-like field is fixed to $\langle \phi \rangle \simeq 2$.}
\label{plot2}
\end{figure}

To obtain a viable model with $\Delta > 1/2$, we need to introduce a shift in the stability correction in the K\"ahler potential:
\begin{equation}
\label{kah2}
K = -3 \ln \left(T + \overline{T} - \frac{|\phi|^2}{3} - \frac{|z|^2}{3} + \frac{|z|^4}{\Lambda_z^2} + \frac{|\phi - \Delta|^4}{\Lambda_\phi^2} \right),
\end{equation}
where the dynamical stabilization now occurs around the shifted VEV of $\langle \phi \rangle = \Delta$ rather than about $\langle \phi \rangle = 0$. However, due to complexity of the model, it cannot be solved analytically, and we analyze it numerically. We consider a concrete example with $\Delta = 2$, and choose the following parameters for our numerical study:
\begin{equation}
\label{num}
g = 0.05315 \simeq \frac{1}{18}, \qquad \Lambda_z^2 = \Lambda_\phi^2 = 0.1.
\end{equation}
If we use Eqs.~(\ref{vev1} - \ref{vev4}) with~(\ref{num}), we find:
\begin{equation}
\langle T \rangle \simeq 1.20, \qquad \langle \phi \rangle \simeq 2, \qquad \langle z \rangle \simeq 0.11, \qquad b \simeq 0.56.
\end{equation}
To find the canonically-normalized field $T_R$, which drives inflation, we use equation~(\ref{canT}), and we find that the VEV of canonically-normalized field is given by $\langle T_R \rangle \simeq 1.07$. In Figs.~\ref{plot1} and \ref{plot2} we plot the Starobinsky-like inflationary potential for a case corresponding to C-2 with $\Delta = 2$. Fig. \ref{plot1} shows the effective scalar potential $V$ as a function of fields $T_R$ and $\phi$. The Starobinsky-like inflationary plot with fixed value of $\langle \phi \rangle \simeq 2$ is shown in Fig.~\ref{plot2}.

Next, we find the relevant scalar and fermion masses for our particular example. The scalar masses are given by:
\begin{subequations}
\begin{align}
m_{T_R} &\simeq m, \\
m_{T_I} &\simeq m, \\
m_{\phi_R} &\simeq 22.59 \, m, \\
m_{\phi_I} &\simeq 21.84 \, m, \\
m_{z_R} &\simeq 49.21 \, m, \\
m_{z_I} &\simeq 49.23 \, m, 
\end{align}
\end{subequations}
where in this case we no longer have maximal mixing between the fields $T_{R}$ and $\phi_R$ or $T_{I}$ and $\phi_I$. If we eliminate the Goldstino mode from the spectrum, which in this case is a mixture of the fermion fields $\chi_T$, $\chi_{\phi}$, and $\chi_z$, given by:
\begin{equation}
\eta \simeq -2.03 \, \chi_T + 2.43 \, \chi_{\phi} + 0.48 \, \chi_z,
\end{equation}
we find the following fermion masses:
\begin{equation}
m_{\chi_{1}} = 2.82 \, m ,
\end{equation}
\begin{equation}
m_{\chi_{2}} = 1.39 \, m.      
\end{equation}
Because of the substantial shifts in the VEVs given in Eqs. (\ref{vev1} - \ref{vev4}) due to large $\Delta$, we can no longer simply approximate $m_{3/2} = m \Delta/\sqrt{3}$. Instead we find numerically that
the gravitino mass is given by:
\begin{equation}
m_{3/2} \simeq 3.94 \, m,
\end{equation}
where as expected for C-2 models, we have $m_{3/2} > m$.

\begin{figure}[!ht]
\centering
\includegraphics[scale=.55]{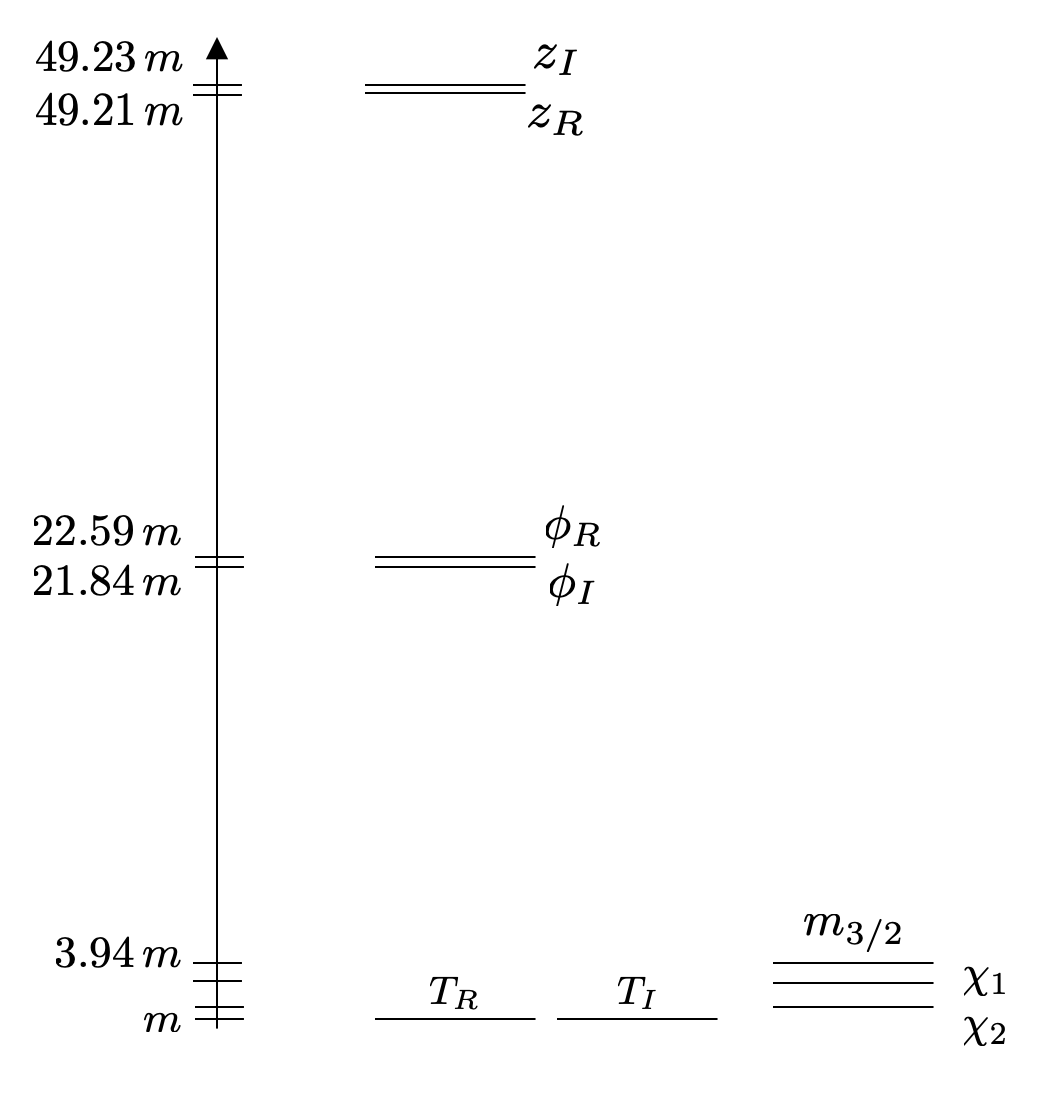}
\caption{Mass spectrum of model C-2. In this scenario, $T_R$ is identified as the inflaton, and $m_{T_R} \simeq m < m_{3/2}$.}
\label{spectrum2}
\end{figure}

Finally, we show the mass spectrum for this particular example of model C-2 in Fig.~\ref{spectrum2}.

\section{Leptogenesis in Cecotti-type Models}
\label{lctype}
In this section we study Cecotti-type models of non-thermal leptogenesis for high-scale supersymmetry. Despite the 
differences in their spectra, leptogenesis in both types of models,
C-1 and C-2, is the same. Of course, in C-1, the gravitino may be the dark matter produced during reheating or direct decays, while in C-2
an additional dark matter candidate must be introduced.

For Cecotti-type models, reheating proceeds via the gravitational coupling of the inflaton $T_R$ to Higgs bosons, and its decay rate is given by~\cite{EGNO4,dgmo,dgkmo,kmo}:
\begin{equation}
\label{decayhiggs}
    \Gamma_{2h} \, = \, \frac{\mu^4}{12 \pi mM_P^2}\equiv\frac{\lambda^2}{8\pi}m,
\end{equation}
where we define a Yukawa-like coupling $\lambda \equiv \sqrt{\frac{2}{3}} \frac{\mu^2}{m \, M_P}$. We also took into account the fact that for high-scale supersymmetry models, the Higgs boson has 4 degrees of freedom. 
The reheating temperature is expressed as~\cite{dgmo}:
\begin{equation}
\label{reh1}
    T_{\rm RH} \simeq \left( \frac{40}{g_* \pi^2}\right) \left( \frac{\Gamma_{2h} \, M_P}{c} \right)^{1/2},
\end{equation}
where $g_* = 427/4$ is the effective degrees of freedom of the Standard model, and $c \simeq 1.2$. We can also express the reheating temperature~(\ref{reh1}) in terms of the coupling $\lambda$ as $T_{\rm RH} \simeq 0.5 (\lambda/2 \pi) \sqrt{m M_P}$, and the maximal temperature attained during reheating is given by~$T_{\text{max}} \simeq 0.5(8 \pi /\lambda^2)^{1/4} \, T_{\rm RH}$.

It was shown in~\cite{dgmo,kmo} that to obtain the correct dark matter relic density, we require the $\mu$-term to be in the range of $m \lesssim \mu \simeq 3 \times 10^{13} - 10^{15} \, \text{GeV}$, which we also expect from the fact the Higgsino mass parameter should lie above the inflaton mass in high-scale supersymmetry models. We then express the reheating temperature~(\ref{reh1}) as:
\begin{equation}
\label{reh2}
    T_{\rm RH} \simeq 7.73 \times 10^{10} \, \text{GeV} \left( \frac{\mu}{10^{14} \, \text{GeV}} \right)^2 \, \left( \frac{m}{3 \times 10^{13} \, \text{GeV}} \right)^{-1/2}.
\end{equation}

The relic density of gravitinos is dependent on the reheat temperature and, through Eq.~(\ref{reh2}), on $\mu$. As noted earlier, there are two contributions to the gravitino relic density. It is produced thermally
by the annihilations of Standard Model particles and directly through 
inflaton decays. The annihilations depend on $T_{\rm RH}^7$ \cite{bcdm,eev,dgmo,dgkmo,kmo} and hence on $\mu^{14}$. Inflaton decays
may also produce a sizeable contribution to the gravitino density.
Tree level decays are suppressed for small $\Lambda_z$ but decays through Higgs loops are always present and in fact dominate over the annihilations when $m_{3/2} \lesssim 0.1 m$ \cite{kmo}. The gravitino abundance through decays depends linearly on $T_{\rm RH} \, \propto \, \mu^2$, which has also a weak (logarithmic) dependence of the branching ratio on $\mu$ \cite{kmo}.

In the case of non-thermal leptogenesis, we must satisfy the constraint $T_{\rm RH} \lesssim M_1$, and if we assume the lower bound of the $\mu$-term, given by $\mu \simeq m \simeq 3 \times 10^{13} \, \text{GeV}$, we obtain:
\begin{equation}
    M_1 \gtrsim 7 \times 10^{9} \, \text{GeV}.
\end{equation}

We begin our analysis by considering the following addition to the  superpotential, which characterizes the seesaw mechanism and leptogenesis in Cecotti-type models:
\begin{equation}
\label{supseesaw1}
W \supset y_{i \alpha} N_i L_{\alpha} H_u + \frac{1}{2} N_i M_i N_i,
\end{equation}
where $i = 1, 2, 3$, $\alpha = e, \mu, \tau$, and $y$ is the Yukawa coupling matrix, where for simplicity we have assumed that the right-handed neutrino mass matrix $M$ is diagonal. We assume the mass hierarchy $2M_1< m \ll M_2, M_3$, and in all cases that we consider the decays of the lightest right-handed neutrino $N_1$ will be responsible for the dominant contribution to the generation of the lepton asymmetry.

Next, we also consider the two-body decay channel of the inflaton to the lightest of the right-handed neutrinos, which can be calculated from the superpotential~(\ref{supseesaw1}), and is given by~\cite{EGNO4}:
\begin{equation}
\label{decayN1}
\Gamma_{2N_1} = \frac{M_1^2 m}{192 \pi M_p^2} \left(1 - \frac{4 M_1^2}{m^2} \right)^{3/2},
\end{equation}
where we included the kinetic factor $(1-4M_1^2/m^2)^{3/2}$ to account for cases when $2M_{1} \lesssim m$. Most importantly, this decay channel will be responsible for the non-thermal production of lightest right-handed neutrinos, which then decay into leptons and Higgs bosons, and produce a lepton asymmetry. Because our inflaton mass is $m \simeq 3 \times 10^{13}~\text{GeV}$ and $\mu \gtrsim m$, the decay channel to Higgs bosons is the dominant channel, and $\Gamma_{2h} \gg \Gamma_{2N_1}$. The branching ratio of the two decay channels is given by:
\begin{equation}
\label{br1}
    B_{R} = \frac{\Gamma_{2N_1}}{\Gamma_{2h}} \simeq \frac{M_1^2 m^2}{16 \mu^4}  \left( 1 - \frac{4 M_1^2}{m^2} \right)^{3/2} \lesssim 10^{-3}.
\end{equation}
It is important to note that we assume that the lightest right-handed neutrino $N_1$ decays instantaneously to leptons and Higgs bosons. As such, we must require $\Gamma_{L_{\alpha} h} > \Gamma_{2N_1}$, which will be justified at the end of this section. 

In order to obtain the number density of the lightest right-handed neutrinos $n_{N_1}$, we assume non-instantaneous reheating. In this case, we find~\cite{kmo}:
\begin{equation}
\label{numb1}
n_{N_1}(T_{RH}) = \frac{g_{*} \, \pi^2}{18 \, m} T_{RH}^4 \, N \, B_R,
\end{equation}
where $N$ is the number of the lightest right-handed neutrinos $N_1$ produced by the inflaton $T_R$ decay, which is $N=2$ in our case, and $B_R$ is the branching ratio of the inflaton to right-handed neutrino decay, given by~(\ref{br1}).
The ratio of the number density of $N_1$ to entropy is:
\begin{equation}
\frac{n_{N_1}}{s} \simeq  \frac{5 m M_1^2 T_{\rm RH}}{32 \mu^4}  \left( 1 - \frac{4 M_1^2}{m^2} \right)^{3/2},
\label{nn1s}
\end{equation}
where the entropy density is $s =  \frac{2 \pi^2}{45} g_{*} T^3$.

\begin{figure}[!ht]
\centering
\includegraphics[scale=.6]{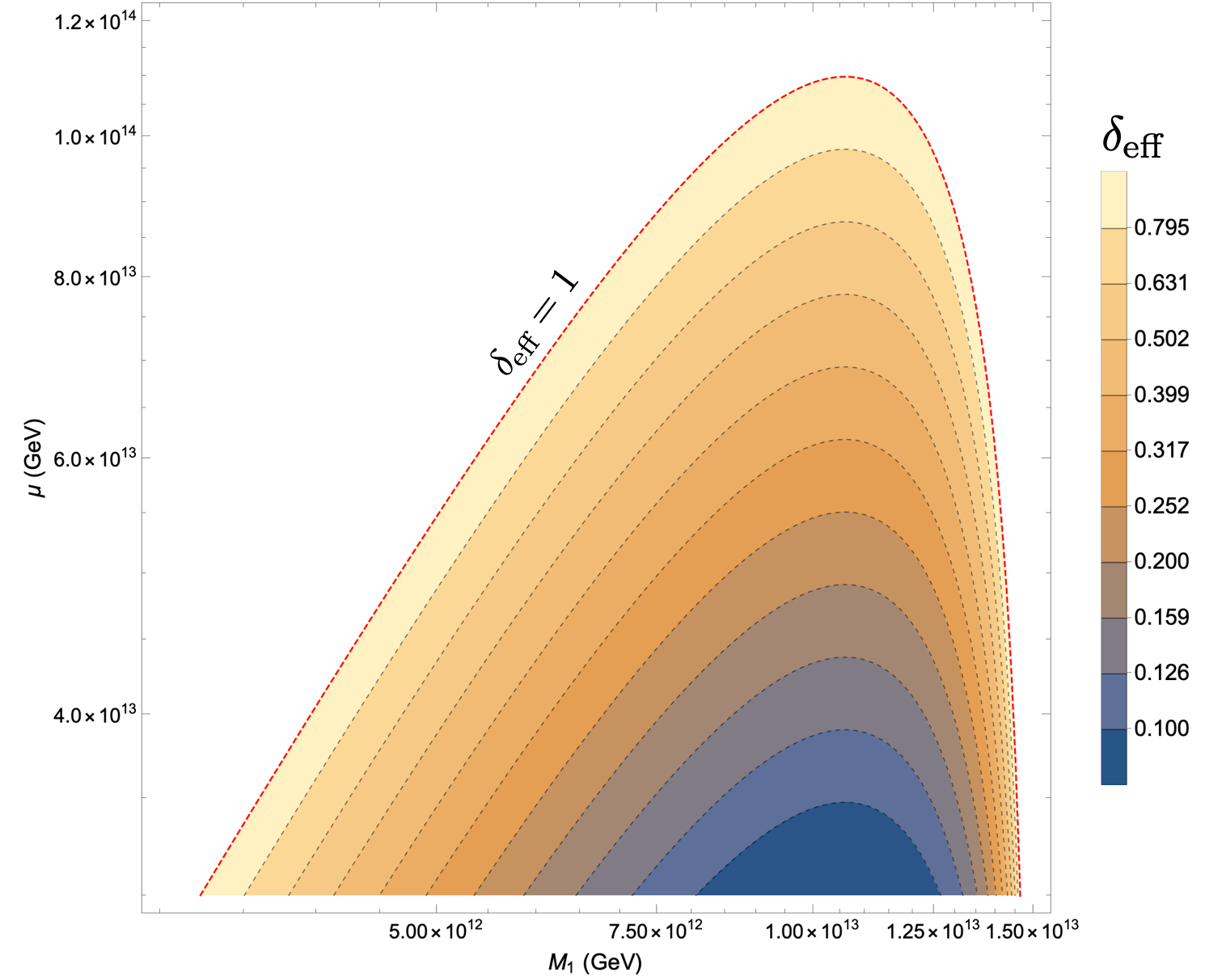}
\caption{Range of right-handed neutrino masses $M_1$ and the $\mu$-term satisfying the baryon asymmetry $Y_B \simeq 8.7 \times 10^{-11}$, with $m_{\nu_3} \simeq 0.05 \, \text{eV}$ in models C-1 or C-2. The red-dashed line corresponds to limit of maximal $CP$-violating phase $\delta_{\text{eff}}$.}
\label{fig:C1}
\end{figure}

\begin{figure}[!ht] 
\centering
\includegraphics[scale=.6]{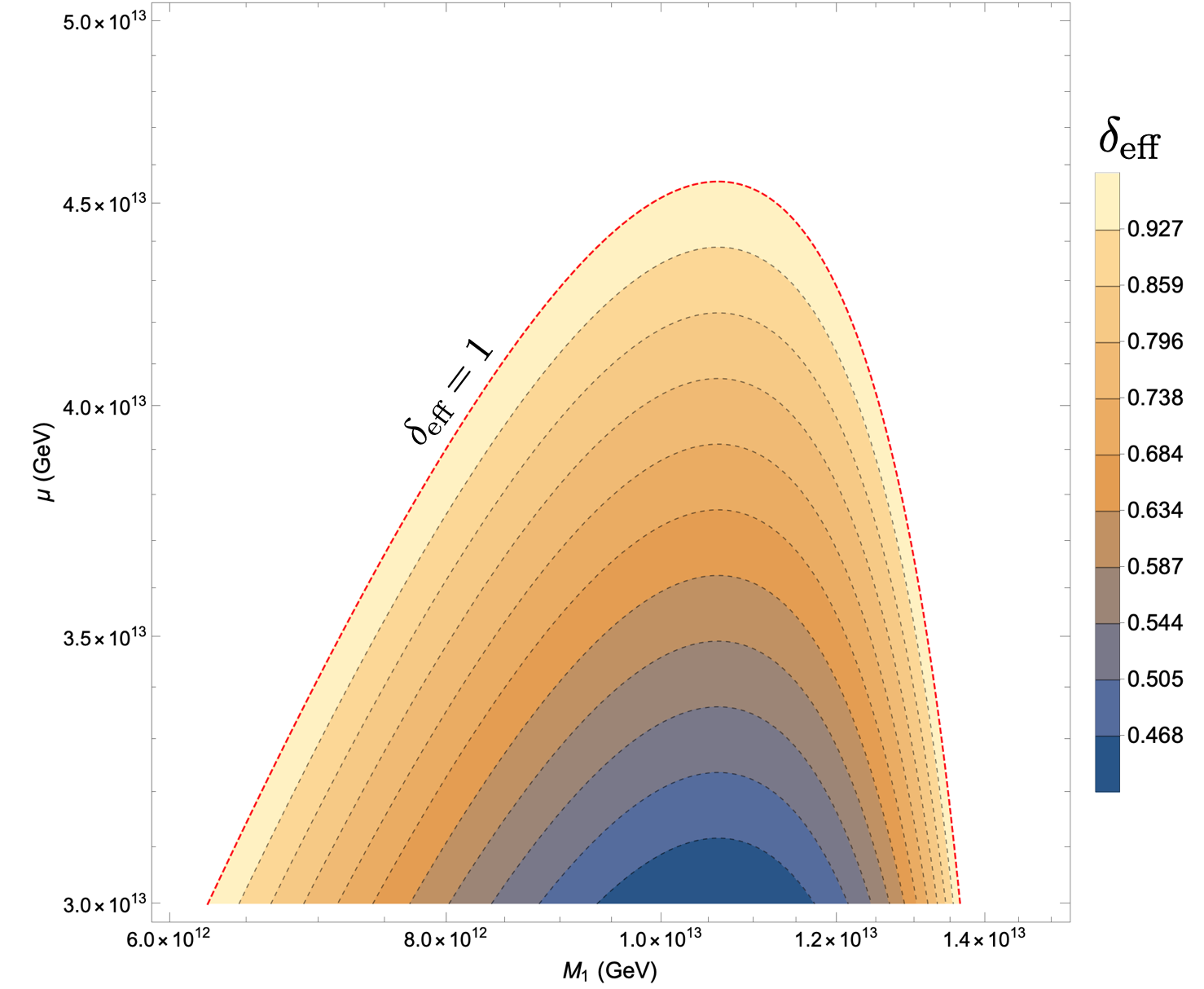}
\caption{Range of right-handed neutrino masses $M_1$ and the $\mu$-term satisfying the baryon asymmetry $Y_B \simeq 8.7 \times 10^{-11}$, with $m_{\nu_2} \simeq 0.0086 \, \text{eV}$  in models C-1 or C-2. The red-dashed line corresponds to limit of maximal $CP$-violating phase $\delta_{\text{eff}}$.}
\label{fig:C2}
\end{figure}

To obtain the baryon asymmetry, we can use Eqs.~(\ref{nn1s}) and (\ref{baryonasy2}) and find:
\begin{small}
\begin{equation}
\label{asy3}
Y_B \simeq 2.5 \times 10^{-13} \delta_{\text{eff}} 
\left(\frac{\mu}{ 10^{14} \, \text{GeV}} \right)^{-2}
\left(\frac{m_{\nu_i}}{0.05 \, \text{eV}} \right)
\left(\frac{M_1}{10^{12} \, \text{GeV}} \right)^{3}
\left( \frac{m}{3 \times 10^{13} \, \text{GeV}} \right)^{1/2}
\left( 1 - \frac{4 M_1^2}{m^2} \right)^{3/2},
\end{equation}
\end{small}
where $i = 2,3$ for the normal hierarchy given by Eq. (\ref{normhier}), and $i = 1,2$ for the inverse hierarchy given by Eq. (\ref{invhier}). Therefore, if we use the observationally determined value for the baryon asymmetry $Y_B \simeq 8.7 \times 10^{-11}$, we obtain the following constraint:
\begin{equation}
\label{rhn1}
\delta_{\text{eff}}^{1/3} \,
\left(\frac{\mu}{ 10^{14} \, \text{GeV}} \right)^{-2/3}
\left(\frac{m_{\nu_i}}{0.05 \, \text{eV}} \right)^{1/3}
\left(\frac{M_1}{10^{12} \, \text{GeV}} \right)
 \left( \frac{m}{3 \times 10^{13} \, \text{GeV}} \right)^{1/6}
\left( 1 - \frac{4 M_1^2}{m^2} \right)^{1/2} \simeq 7 .
\end{equation}
This constraint is used in Fig.~\ref{fig:C1} to find the allowed values $\mu$, $M_1$, and $\delta_{\text{eff}}$, that can accommodate the observed value of baryon asymmetry $Y_B$ for fixed $m = 3 \times 10^{13}$ GeV and $m_{\nu_3} \simeq 0.05 \, \text{eV}$.
For each pair of points ($M_1, \mu$), the shading corresponds to the required value of $\delta_{\rm eff}$ to obtain the correct 
baryon asymmetry.  An analogous
plot using $m_{\nu_2} \simeq 0.0086 \, \text{eV}$ is shown in Fig.~\ref{fig:C2}.

We see from these figures that there is a maximum value of 
$\mu \lesssim 1.1 \times 10^{14}$ GeV ($4.6 \times 10^{13}$ GeV)
using $m_{\nu_3}$ ($m_{\nu_2}$) in Eq. (\ref{rhn1}).  It is rather amazing that the range of $\mu$ required to obtain the correct baryon asymmetry coincides with the value of $\mu$ needed to obtain the correct relic density of gravitino dark matter in C-1 type models (recall, there is no gravitino dark matter in C-2 models). 

In the case of the inverted hierarchy of the left-handed neutrinos, using Eq. (\ref{invhier}), the limits in the ($M_1, \mu, \delta_{\rm eff}$) parameter space are very similar to the results shown in Fig.~(\ref{fig:C1}) because \textit{IH} neutrino masses are very close to $m_{\nu_3} \simeq 0.05 \, \text{eV}$ in the \textit{NH}.

In order to ensure that the lightest right-handed neutrino decays immediately out-of-equilibrium, we need to satisfy the constraint $\Gamma_{L_{\alpha} h} > \Gamma_{2N_1}$. The decay rate of the lightest right-handed neutrino $N_1$ is given by:
\begin{equation}
\label{decayN2}
\Gamma_{L_{\alpha} h} = \frac{(yy^{\dagger})_{11}}{4 \pi} M_1 \equiv \frac{|y_1|^2}{4 \pi} M_1  = \frac{m_{\nu_1} \, M_1^2}{4 \pi \, v^2 \sin^2 \beta}.
\end{equation}
where we have included decays to $L_\alpha h$ and ${\bar L}_\alpha \bar{h}$ and we denoted the dominant contribution of $(yy^{\dagger})_{11} \simeq |y_1|^2$. If we now compare the decay rate~(\ref{decayN2}) to the inflaton decay rate into the right-handed neutrinos~(\ref{decayN1}), we find:
\begin{equation}
\label{neut1cons}
m_{\nu_1} \gtrsim 1.6 \times 10^{-12} \, \text{eV},
\end{equation}
which is clearly easily satisfied.

\section{Wess-Zumino-type Models of Inflation}
\label{wz-type}

In this section, we consider models based on the Wess-Zumino superpotential (\ref{WZpot}),  supersymmetry breaking 
is most easily attained by simply adding a constant, ${\tilde m}$,
to the superpotential, giving $m_{3/2} = {\tilde m}$. Indeed if the constant is promoted
to a Polonyi term in $W$, then the inflationary potential
is affected and it becomes difficult to maintain a flat potential at large field values~\cite{EGNO4,king,dgmo}
with high-scale supersymmetry breaking. 

In the absence of a Polonyi term, using only a constant term in $W$, supersymmetry breaking is
generated by an $F$-term for $T$. Gaugino masses are given
by Eq. (\ref{gaugino}) upon replacing $z \to T$.
However, in this case, because the VEV of $T$ is of order
the Planck scale $M_P$, even if we write $f = f_0 + f_1 T/\Lambda_T$, if $\Lambda_T \ll \langle T \rangle$,
we are inevitably led to $m_{1/2} \sim m_{3/2}$
since we must require $f_1 \langle T \rangle/\Lambda_T \la (1/g^2)$. Thus there is no simple way to realize
a high-scale supersymmetry model with
$m_{1/2} \sim m_0 > m > m_{3/2}$. Therefore, we no longer consider WZ-1 models with $m_{3/2} < m$. 

For WZ-2 models with $m_{3/2} > m$, the above problems are no longer present, as we can again break supersymmetry
with a constant superpotential term, with $m_{1/2} \sim m_0 \sim m_{3/2} > m$. The constant term $\tilde{m}$, which breaks supersymmetry, does not shift the minimum, and at the end of inflation we are left with $\langle T_R \rangle = 1/2$, $\langle T_I \rangle = 0$, and $\langle \phi_R \rangle = \langle \phi_I \rangle = 0$. As in the case of C-2 models, we can no longer consider the gravitino as a dark matter candidate. 

The mass spectrum for the Wess-Zumino type models is relatively simple. Both the real and imaginary parts of the inflaton
have a common mass:
\begin{equation}
    m_{\phi_{R,I}}^2 \simeq m^2 .
\end{equation}
In order to ensure the stability of the potential during inflation, the $T$ field is dynamically stabilized with the higher-order terms in Eq.~(\ref{K2}). As in the case of the strongly stabilized Polonyi field, the stabilization of volume modulus $T$ results in a mass, which is hierarchically higher than the gravitino mass: 
\begin{equation}
    m_{T_{R}}^2 \simeq \frac{48 m_{3/2}^2}{\Lambda_{T}^2}, \qquad m_{T_{I}}^2 \simeq \frac{48 \, d^2 m_{3/2}^2}{\Lambda_{T}^2},
\end{equation}
where the constant $d$ was define in Eq. (\ref{d}).
As we start with only 2 superfields in this case, and supersymmetry
is broken, there is only one chiral fermion which is associated with the inflaton. Namely, the inflatino mass is simply:
\begin{equation}
    m_{\chi_{\phi}} \simeq m . 
\end{equation}
The fermion associated with $T$ is the Goldstino and becomes
the longitudinal component of the gravitino and $m_{3/2} = {\tilde m}$. 
The mass spectrum  for WZ-2 is illustrated in Fig. \ref{spectrum3} below.
\begin{figure}[!ht]
\centering
\includegraphics[scale=.6]{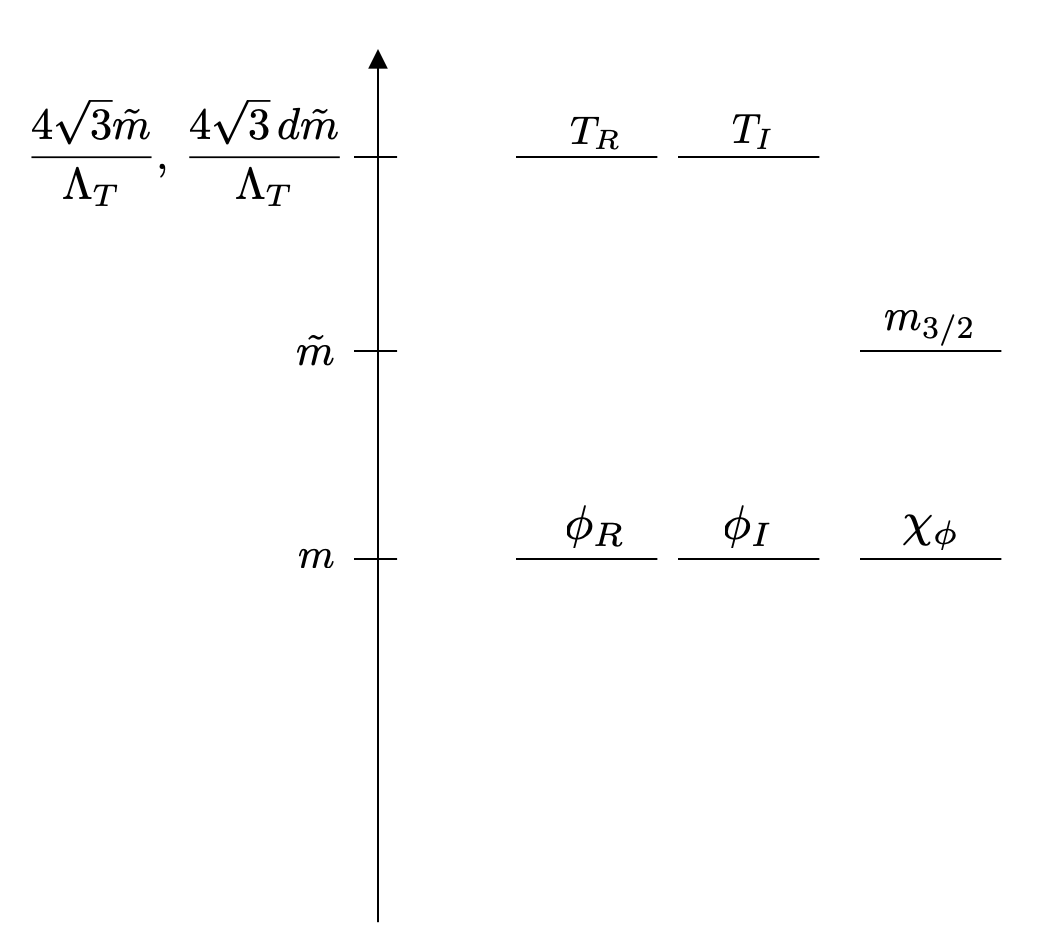}
\caption{Mass spectrum of model WZ-2. In this scenario, $\phi_R$ is identified as the inflaton, and $m_{\phi_R} \simeq m < m_{3/2}$. }
\label{spectrum3}
\end{figure}

\section{Leptogenesis in Wess-Zumino-type Models}
\label{lwztype}

A key difference between Cecotti and Wess-Zumino models of 
inflation in no-scale supergravity, is the manner in which the Universe reheats \cite{EGNO4}. In models where the inflaton is associated the volume modulus, $T$, the inflaton couples
to Standard Model fields (and their supersymmetric partners).
Thus there are many open decay channels leading to reheating.
In high-scale supersymmetry, as discussed earlier, the dominant
decay mode is the two-body decay to two Higgs bosons. However,
in Wess-Zumino models, where the inflaton is associated with 
a matter-like field $\phi$, in the absence of a direct coupling of the inflaton to Standard Model fields, there are no decay channels available \cite{ekoty,EGNO4}. Reheating in this case typically 
relies on a coupling of the inflaton in the gauge kinetic function 
which then allows for decays to gauge bosons (and in the case of low scale supersymmetry, to gauginos) \cite{ekoty,klor,EGNO4}.

In Wess-Zumino-like models, it is, however, possible to associate 
the inflaton with a right-handed sneutrino \cite{eno8}~\footnote{The association of the matter-like field $\phi$ with $N$ is not possible in Cecotti-like models. When supersymmetry is broken, $\phi$ gets a VEV given by Eq. (\ref{cshiftsf}), which approaches the Planck scale in high-scale supersymmetric models. This VEV induces a bilinear R-parity violating term which induces gravitino decay, and is strongly constrained in order for the gravitino lifetime to remain sufficiently long \cite{dgkmo,dgkmo2}. This bound translates into $y_\nu \lesssim 10^{-21}$, thus preventing it in generating a neutrino mass and lepton asymmetry.}. For example, starting with with 
the Wess-Zumino superpotential~(\ref{WZpot}) 
we can equate $\phi$ with $N_2$ (or $N_3$, however, we still require $2 M_1 \lesssim m$ for leptogenesis, so $\phi$ can not be $N_1$). 
Thus, we consider the superpotential
\begin{equation}
\label{supwz}
W \supset - m \left(\frac{N^3_2}{3 \sqrt{3}} \right) + y_{i \alpha} N_i L_{\alpha} H_u + \frac{1}{2} N_i M_i N_i \, ,
\end{equation}
where $i=1,2,3$.
From Eq. (\ref{supwz}), we see that there is a direct coupling
of the inflaton to $L_\alpha$ and $H_u$ with Yukawa coupling $y_{2\alpha}$. To preserve the form of the Starobinsky potential,
we must require $M_2 = m$. For leptogenesis, we require a decay
of the inflaton to $N_1$ and
assume the mass hierarchy $2M_1 \lesssim m < M_{3}$.

In models of weak scale supersymmetry, the Yukawa coupling of the inflaton to $L_\alpha H_u$ leads to efficient reheating after inflation \cite{eno8}. However, in high-scale supersymmetry with 
${\tilde m} > m$, the two possible tree level decays (slepton + Higgs, or lepton + Higgsino) are both kinematically forbidden. 
One loop decays to Standard Model fields are possible, but these are suppressed. It is, however, possible to introduce a superpotential coupling 
\begin{equation}
W \supset - \frac12 \kappa \, {N}_2 N_1 N_1  \, ,
\end{equation}
which leads to the following trilinear term in the Lagrangian,  
\begin{equation}
\mathcal{L} \supset - \kappa \, \tilde{N}_2 N_1 N_1 + \text{h.c.} \, ,
\end{equation}
which leads to the following decay rate for the inflaton to two lighter right-handed neutrinos:
\begin{equation}
\label{decaywz1}
\Gamma_{2 N_1} = \frac{\kappa^2 m}{8 \pi} \left(1-\frac{4M_1^2}{m^2} \right)^{3/2}
\end{equation}
This decay to 
two right-handed neutrinos, $N_1$, dominates and is responsible for
generating reheating and leptogenesis.  The reheat temperature is given by Eq. (\ref{reh1}) with the substitution $\Gamma_{2h} \to \Gamma_{2{N_1}}$.
We can express the reheating temperature as:
\begin{equation}
\label{rehwz2}
T_{RH} \simeq 6.8 \times 10^{14}  \, \kappa \, \text{GeV} \left( \frac{m}{3 \times 10^{13} \, \text{GeV}}\right)^{1/2} \left(1-\frac{4M_1^2}{m^2} \right)^{3/4}.
\end{equation}
Assuming non-instantaneous reheating, the number density of $n_{N1}$ is again given by Eq.~(\ref{numb1}) with $N=2$.
We find that the number density to entropy ratio is
\begin{equation}
\frac{n_{N_1}}{s} = \frac{5 T_{\rm RH}}{2 m} \simeq 57 \, \kappa \, \left( \frac{m}{3 \times 10^{13} \, \text{GeV}}\right)^{-1/2} \left(1-\frac{4M_1^2}{m^2} \right)^{3/4}
\end{equation}
and we can express the baryon asymmetry~(\ref{baryonasy2}) as:
\begin{equation}
\label{wzasy}
Y_B \simeq 4 \times 10^{-3} \delta_{\text{eff}} \, \, \kappa \, \left( \frac{m_{\nu_i}}{0.05 \, \text{eV}} \right) \left( \frac{M_1}{10^{12} \, \text{GeV}} \right) 
\left( \frac{m}{3 \times 10^{13} \, \text{GeV}}\right)^{-1/2}
\left(1-\frac{4M_1^2}{m^2} \right)^{3/4}.
\end{equation}
If we connect it to the observed baryon asymmetry of the Universe, $Y_B \simeq 8.7 \times 10^{-11}$, we obtain the following constraint:
\begin{equation}
    \delta_{\text{eff}} \, \, \kappa \, \left( \frac{m_{\nu_i}}{0.05 \, \text{eV}} \right) \left( \frac{M_1}{10^{12} \, \text{GeV}} \right)
    \left( \frac{m}{3 \times 10^{13} \, \text{GeV}}\right)^{-1/2}
    \left(1-\frac{4M_1^2}{m^2} \right)^{3/4} \simeq 2.2 \times 10^{-8}.
    \label{wzcons}
\end{equation}

\begin{figure}[!ht]
\centering
\includegraphics[scale=.55]{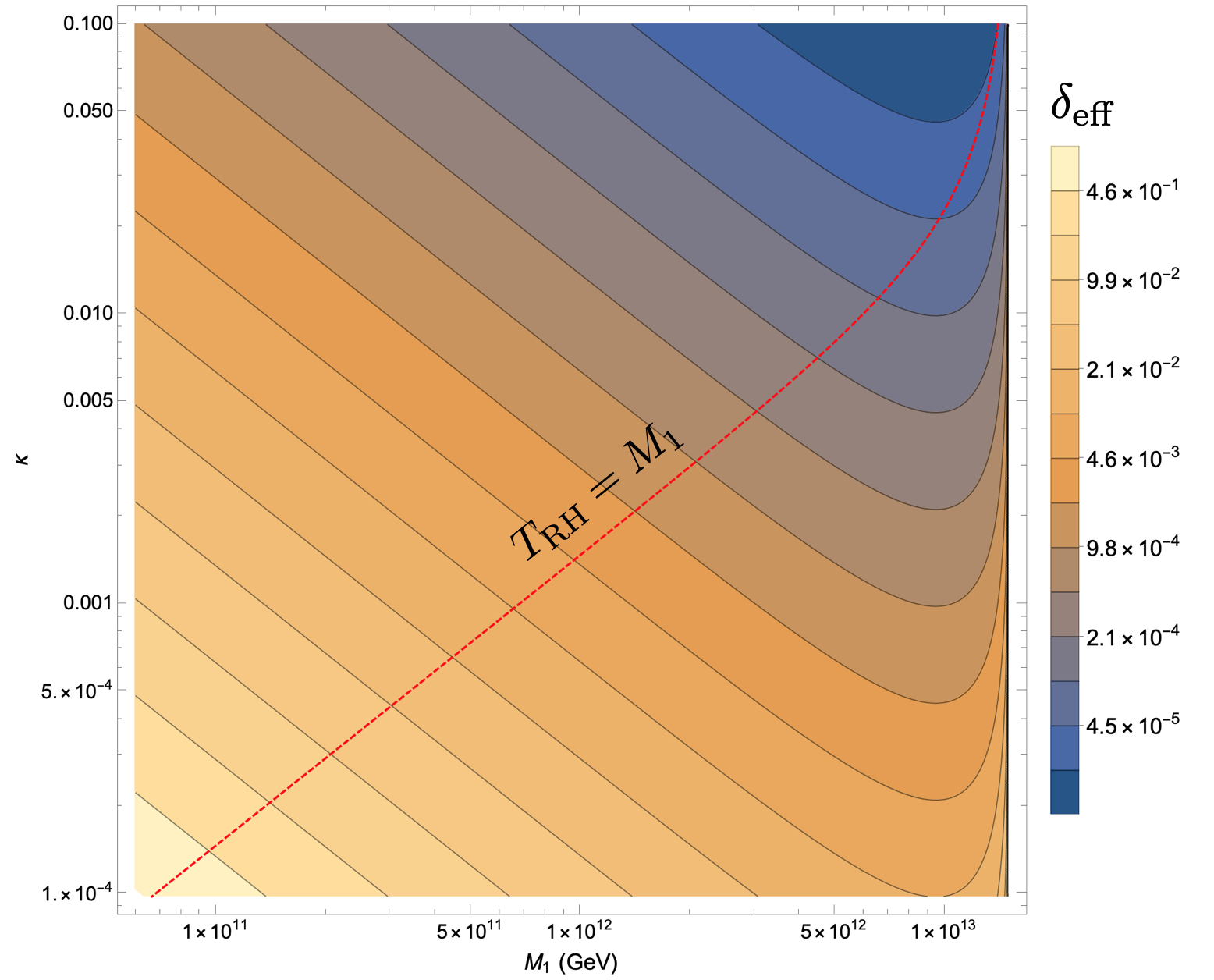}
\caption{$CP$-violating effective phase $\delta_{\text{eff}}$ as a function of the lightest right-handed neutrino mass $M_1$ and the trilinear coupling $\kappa$ for $m_{\nu_3} \simeq 0.05 \, \text{eV}$ for model WZ-2. The area below the red-dashed line shows the region when $T_{\rm RH} \lesssim M_1$, which is necessary for non-thermal leptogenesis.}
\label{fig:WZ22}
\end{figure}

\begin{figure}[!ht]
\centering
\includegraphics[scale=.55]{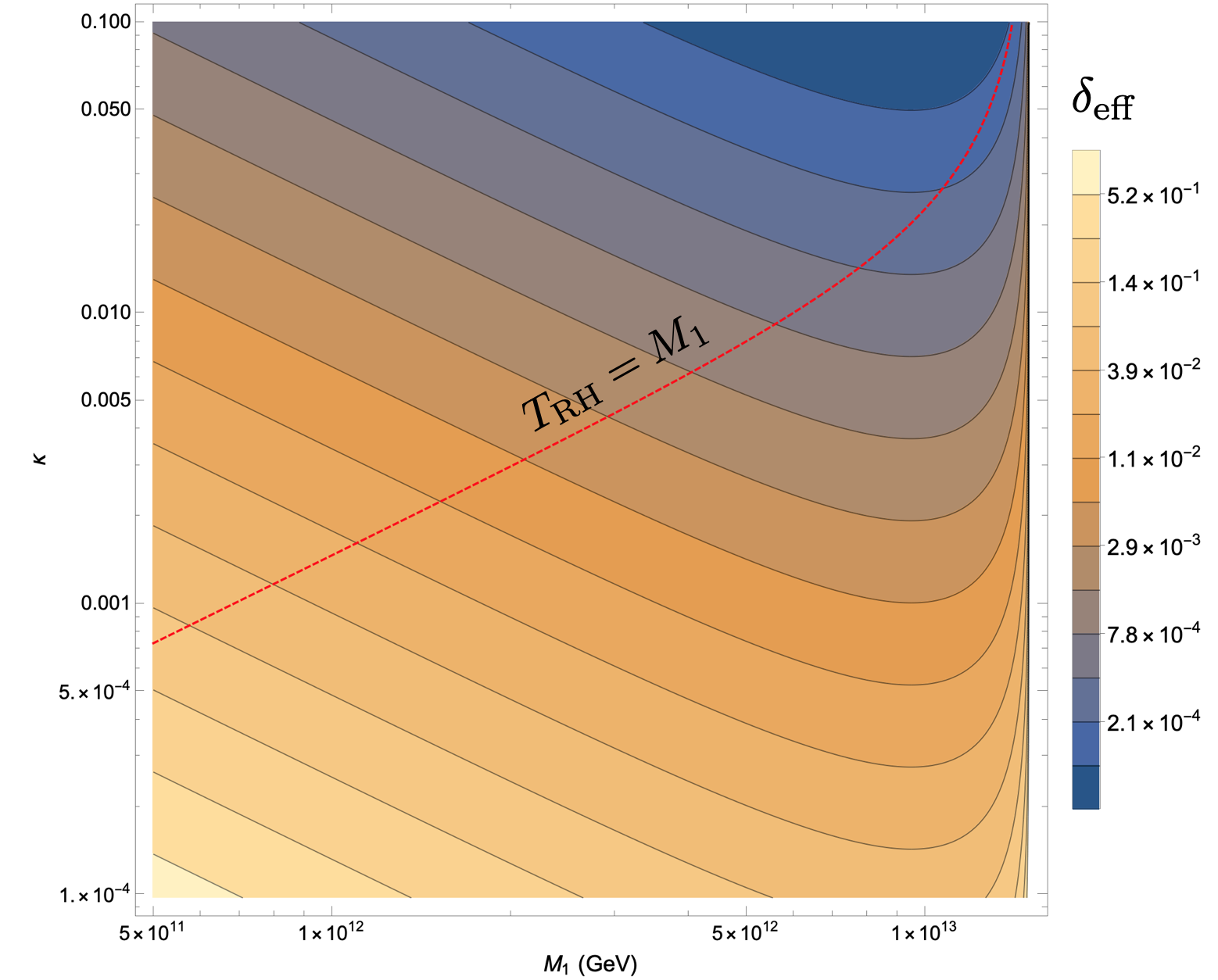}
\caption{$CP$-violating effective phase $\delta_{\text{eff}}$ as a function of the lightest right-handed neutrino mass $M_1$ and the Yukawa coupling $\kappa$ for $m_{\nu_2} \simeq 0.0086 \, \text{eV}$ for model WZ-2. The area below the red-dashed line shows the region when $T_{\rm RH} \lesssim M_1$, which is necessary for non-thermal leptogenesis.}
\label{fig:WZ23}
\end{figure}

For fixed $m_\nu$ and $m$, the constraint in Eq.~(\ref{wzcons}) on the ($\kappa, M_1, \delta_{\rm eff}$) parameter space 
is plotted in Figs.~\ref{fig:WZ22} and \ref{fig:WZ23}.
For each pair ($M_1,\kappa$), the shading determines the required
value of $\delta_{\rm eff}$ needed to obtain the correct baryon asymmetry. 
Also plotted is the boundary for which $T_{\rm RH} = M_1$. For
values of $\kappa$ above this line, $T_{\rm RH} > M_1$, and one
must consider thermal leptogenesis \cite{thermlept}. As one can see from the figures, this is a relatively efficient mechanism for generating the baryon asymmetry, though it does not require
particularly small couplings or phases. 


Finally, we compare the decay rate of the lightest right-handed neutrino $N_1$ with the inflaton decay rate as a check on its out-of-equilibrium decay. 
The decay rate for $N_1$ is again given by Eq.~(\ref{decayN2})
and must be compared with the inflaton decay rate in Eq.~(\ref{decaywz1}). Requiring $\Gamma_{L_{\alpha}h} \gtrsim \Gamma_{2N_1}$ we find:
\begin{equation}
\label{yukcons1}
\frac{m_{\nu_1} \, M_1^2}{\langle H_u \rangle^2} \gtrsim \frac12 \kappa^2 m \left(1-\frac{4M_1^2}{m^2} \right)^{3/2},
\end{equation}
or
\begin{equation}
\label{yukcons2}
m_{\nu_1} \gtrsim 230 \, \kappa^2 \, \text{eV} \left(\frac{M_1}{10^{12} \, \text{GeV}} \right)^{-2} \left(1-\frac{4M_1^2}{m^2} \right)^{3/2}.
\end{equation}
The strongest bound on $m_{\nu_1}$ is found when $\kappa$ is set 
at its maximum value which occurs when $T_{\rm RH} = M_1$ or when
$M_1 = 6.8 \times 10^{14} \kappa$ GeV. In this case, $m_{\nu_1} \gtrsim 5 \times 10^{-4}$ eV. For fixed $M_1$, values of $\kappa$ lower than the
value needed for $T_{\rm RH} = M_1$ reduce this lower bound on $m_{\nu_1}$. This limit is always satisfied for the inverted neutrino hierarchy.

\section{Conclusions}
\label{summary}

Viable models of the very early Universe must account for both inflation
and the generation of net baryon asymmetry.  
In the models presented here, although both of these mechanisms were operative near the scale of grand unification, neither are explicitly dependent on a
specific model of grand unification. Furthermore, while these models 
are inherently supersymmetric, supersymmetry is broken at a scale above the inflationary scale of $m = 3 \times 10^{13}$ GeV and aside from the possible 
role of the gravitino (if $m_{3/2} < m$), supersymmetry does not affect the low
energy behavior of the theory.  

The models of inflation we consider are based on no-scale supergravity.
The inflationary sector requires two chiral superfields \cite{Avatars}
which parametrizes a non-compact $\frac{SU(2,1)}{SU(2) \times U(1)}$ coset
manifold. In a particular basis, one of the fields may be associated with 
the volume modulus while the second appears as a matter-like field. 
Due to the underlying symmetry of the theory, either of the fields can
play the role of the inflaton \cite{ENO6,Avatars,ENOV1} whose potential 
can take the form of the Starobinsky potential \cite{Staro}.
However once supersymmetry is broken, and couplings to the 
Standard Model are introduced, these two classes of inflationary
models appear quite different. 

Supersymmetry breaking can be achieved by simply adding a constant term to the superpotential \cite{EGNO4,enno}, by introducing a Polonyi sector \cite{pol}, or in the case of a matter-like inflaton, by adding a linear term to the superpotential \cite{king}. For modulus-driven inflation, adding a constant
to the superpotential perturbs the potential in such a way so as to always
lead to an AdS vacuum \cite{EGNO4}. Adding a Polonyi sector, 
preserves the form of the potential suitable for inflation.
This is true independent of the scale of supersymmetry breaking ${\tilde m}$, although as we have shown, for ${\tilde m} > m$, additional superpotential couplings are needed. In the case of a matter-like inflaton, a linear term or
a Polonyi sector severely perturbs the inflationary potential for high-scale supersymmetry breaking. In contrast, in this case, adding a constant
term allows an arbitrarily high supersymmetry breaking scale. 

Reheating in the two classes of inflationary models is also quite
different. The volume modulus couples to all sectors of the theory,
and in the case of high-scale supersymmetry breaking, final state 
Higgs bosons are the dominant
decay mode \cite{EGNO4,dgmo,dgkmo,kmo}. In contrast, without an explicit
superpotential coupling to the Standard Model, reheating for a matter-like inflaton
occurs only if inflaton couples to the gauge sector through the gauge kinetic function \cite{ekoty,klor,EGNO4}. As a consequence, we have here associated
the inflaton with one of the right-handed sneutrinos. 

Subsequent to reheating, we have considered in detail
mechanisms for leptogenesis \cite{FY}. Assuming the existence of a right-handed neutrino sector, we have assumed that one of the Majorana right-handed neutrinos are lighter than the inflationary scale, $M_1 \lesssim m$.
In the case of modulus-driven inflation, the branching ratio to right-handed
neutrinos from inflaton decay is calculated. Once produced during reheating, the right-handed neutrinos decay quickly (their decay rate is faster than their
production rate so long as the lightest left-handed neutrino has $m_{\nu_1} >
3.3 \times 10^{-12}$ eV) and decay out-of-equilibrium if $T_{\rm RH} < M_1$. 
The resulting lepton asymmetry is then converted to a baryon asymmetry
via sphaleron interactions.

In the case of a matter-like inflaton, a coupling of the inflaton (one of the heavier two right-handed sneutrinos) decays predominantly into the lightest right-handed neutrino, and its out-of-equilibrium decay simultaneously reheats the Universe and produces the lepton asymmetry.
In this case, the departure from thermal equilibrium requires 
 $m_{\nu_1} > 10^{-3}$ eV.
 
 In both cases of a modulus-like and matter-like inflaton,
 we distinguish between a supersymmetry breaking scale in which the 
 gravitino mass is above or below the inflationary scale.
 In both cases, we have derived the mass spectra of the inflationary/SUSY breaking sectors. 
 When $m_{3/2} < m$ the gravitino may be the dark matter \cite{eev}. 
 A hierarchy between the gravitino mass and the supersymmetry breaking
 scale is possible with the aid of a strongly stabilized Polonyi sector
 \cite{dine}. Thus it is only possible for modulus-driven inflation.
 When  $m_{3/2} > m$, both types of inflationary models are viable,
 but a new dark matter candidate is needed.  Integrating such
 a candidate will be the subject of future work.

\section*{Acknowledgements}
This  work was supported by the France-US PICS MicroDark and by Institut Pascal at Université Paris-Saclay with the support of the P2I and SPU research departments and  the P2IO Laboratory of Excellence (program “Investissements d’avenir” ANR-11-IDEX-0003-01 Paris-Saclay and ANR-10-LABX-0038), as well as the IPhT. Y.M. acknowledges partial support from the European Union Horizon 2020
research and innovation programme under the Marie Sklodowska-Curie: RISE
InvisiblesPlus (grant agreement No 690575), the ITN Elusives (grant
agreement No 674896) and the Red Consolider  MultiDark 
FPA2017-90566-REDC.  The work of K.K.,  K.A.O., and S.V. was supported in
part by the DOE grant DE--SC0011842 at the University of Minnesota.
 K.A.O.
 acknowledges support by the Director, Office of Science, Office of High Energy Physics of the U.S. Department of Energy under the Contract No. DE-AC02-05CH11231.
K.A.O. would also like to thank the Department of Physics and the 
high energy theory group
at the University of California, Berkeley as well as the theory group at LBNL
for their hospitality and financial support while
finishing this work.


\begin{thebibliography}{99}

\bibitem{Maiani:1979cx}
L.~Maiani,
in Proceedings, Gif-sur-Yvette Summer School On Particle Physics,
  1979, 1-52;
Gerard 't~Hooft and others (eds.),
{\it Recent Developments in Gauge Theories, Proceedings of the Nato Advanced
  Study Institute, Cargese, France, August 26 - September 8, 1979},
Plenum press, New York, USA, 1980, Nato Advanced Study Institutes
  Series: Series B, Physics, 59.;
Edward Witten,
{\em Phys. Lett.} B105, 267, 1981.

\bibitem{Ellis:1990zq}
John~R. Ellis, S.~Kelley and Dimitri~V. Nanopoulos,
{\em Phys. Lett.} B249, 441, 1990;
John~R. Ellis, S.~Kelley and Dimitri~V. Nanopoulos,
{\em Phys. Lett.} B260, 131, 1991;
Ugo Amaldi, Wim de~Boer, and Hermann Furstenau.
\newblock {\em Phys. Lett.}, B260, 447, 1991;
Paul Langacker and Ming-xing Luo,
{\em Phys. Rev.} D44, 817, 1991;
C.~Giunti, C.~W. Kim and U.~W. Lee,
{\em Mod. Phys. Lett.} A6, 1745, 1991.


  \bibitem{Ellis:2000ig} 
J.~R.~Ellis and D.~Ross,
  Phys.\ Lett.\ B {\bf 506}, 331 (2001)
  [hep-ph/0012067].

\bibitem{ewsb}
L.~E.~Ibanez and G.~G.~Ross,
  Phys.\ Lett.\ B {\bf 110}, 215 (1982);
  K.~Inoue, A.~Kakuto, H.~Komatsu and S.~Takeshita,
  Prog.\ Theor.\ Phys.\  {\bf 68}, 927 (1982)
  [Erratum-ibid.\  {\bf 70}, 330 (1983)]
  [Prog.\ Theor.\ Phys.\  {\bf 70}, 330 (1983)];
L.~E.~Ibanez,
  Phys.\ Lett.\ B {\bf 118}, 73 (1982);
  J.~R.~Ellis, D.~V.~Nanopoulos and K.~Tamvakis,
  Phys.\ Lett.\ B {\bf 121}, 123 (1983);
J.~R.~Ellis, J.~S.~Hagelin, D.~V.~Nanopoulos and K.~Tamvakis,
  Phys.\ Lett.\ B {\bf 125}, 275 (1983);
   L.~Alvarez-Gaume, J.~Polchinski and M.~B.~Wise,
  Nucl.\ Phys.\ B {\bf 221}, 495 (1983).

  \bibitem{ehnos}        
  		H.~Goldberg,
                Phys.\ Rev.\ Lett.\ {\bf 50} (1983) 1419;
                J.~Ellis, J.~Hagelin, D.~Nanopoulos, K.~Olive and M.~Srednicki,
                Nucl.\ Phys.\ B {\bf 238} (1984) 453.

 \bibitem{cries}
J.~R.~Ellis, D.~V.~Nanopoulos, K.~A.~Olive and K.~Tamvakis,
  Phys.\ Lett.\  {\bf 118B} (1982) 335;
  J.~R.~Ellis, D.~V.~Nanopoulos, K.~A.~Olive and K.~Tamvakis,
  Nucl.\ Phys.\ B {\bf 221} (1983) 52;
  K.~Nakayama and F.~Takahashi,
  JCAP {\bf 1110}, 033 (2011)
  [arXiv:1108.0070 [hep-ph]].
  
   \bibitem{nost}
   D.~V.~Nanopoulos, K.~A.~Olive, M.~Srednicki and K.~Tamvakis,
  Phys.\ Lett.\ B {\bf 123}, 41 (1983);
     R.~Holman, P.~Ramond and G.~G.~Ross,
  Phys.\ Lett.\ B {\bf 137}, 343 (1984);
   A.~B.~Goncharov and A.~D.~Linde,
  Phys.\ Lett.\ B {\bf 139}, 27 (1984).

 \bibitem{eta}
  E.~J.~Copeland, A.~R.~Liddle, D.~H.~Lyth, E.~D.~Stewart and D.~Wands,
  Phys.\ Rev.\ D {\bf 49}, 6410 (1994)
  [astro-ph/9401011];
E.~D.~Stewart,
  Phys.\ Rev.\ D {\bf 51}, 6847 (1995)
  [hep-ph/9405389].
  
    \bibitem{no-scale}
E.~Cremmer, S.~Ferrara, C.~Kounnas and D.~V.~Nanopoulos,
  Phys.\ Lett.\ B {\bf 133} (1983) 61;
  J.~R.~Ellis, A.~B.~Lahanas, D.~V.~Nanopoulos and K.~Tamvakis,
  Phys.\ Lett.\ B {\bf 134} (1984) 429;
 A.~B.~Lahanas and D.~V.~Nanopoulos,
  Phys.\ Rept.\  {\bf 145} (1987) 1.
  
  \bibitem{nsi}
   A.~S.~Goncharov and A.~D.~Linde,
  Class.\ Quant.\ Grav.\  {\bf 1},  L75 (1984);
C.~Kounnas and M.~Quiros,
  Phys.\ Lett.\ B {\bf 151}, 189 (1985);
J.~R.~Ellis, K.~Enqvist, D.~V.~Nanopoulos, K.~A.~Olive and M.~Srednicki,
  Phys.\ Lett.\  {\bf 152B} (1985) 175
   Erratum: [Phys.\ Lett.\  {\bf 156B} (1985) 452].

  \bibitem{Staro}
A.~A.~Starobinsky,
  Phys.\ Lett.\ B {\bf 91}, 99 (1980).
  
   \bibitem{MukhChib}
    V.~F.~Mukhanov and G.~V.~Chibisov,
  JETP Lett.\  {\bf 33}, 532 (1981)
  [Pisma Zh.\ Eksp.\ Teor.\ Fiz.\  {\bf 33}, 549 (1981)].

\bibitem{planck18}
  N.~Aghanim {\it et al.} [Planck Collaboration],
  arXiv:1807.06209 [astro-ph.CO];
  Y.~Akrami {\it et al.} [Planck Collaboration],
  arXiv:1807.06211 [astro-ph.CO].

  \bibitem{ENO6}
   J.~Ellis, D.~V.~Nanopoulos and K.~A.~Olive,
  Phys.\ Rev.\ Lett.\  {\bf 111} (2013) 111301 
  [arXiv:1305.1247 [hep-th]].

  \bibitem{Avatars}
   J.~Ellis, D.~V.~Nanopoulos and K.~A.~Olive,
  JCAP {\bf 1310} (2013) 009
 [arXiv:1307.3537 [hep-th]].
  
  \bibitem{eno8}
     J.~Ellis, D.~V.~Nanopoulos and K.~A.~Olive,
  Phys.\ Rev.\ D {\bf 89} (2014) 4,  043502
  [arXiv:1310.4770 [hep-ph]];

  \bibitem{eno9}
  J.~Ellis, D.~V.~Nanopoulos and K.~A.~Olive,
  Phys.\ Rev.\ D {\bf 97}, no. 4, 043530 (2018)
  [arXiv:1711.11051 [hep-th]].

  \bibitem{ENOV1}
  J.~Ellis, D.~V.~Nanopoulos, K.~A.~Olive and S.~Verner,
JHEP {\bf 1903}, 099 (2019).
[arXiv:1812.02192 [hep-th]].

\bibitem{ENOV2}
   J.~Ellis, D.~V.~Nanopoulos, K.~A.~Olive and S.~Verner,
  Phys.\ Rev.\ D {\bf 100}, no. 2, 025009 (2019)
  [arXiv:1903.05267 [hep-ph]].
  
  \bibitem{ENOV3}
  J.~Ellis, D.~V.~Nanopoulos, K.~A.~Olive and S.~Verner,
  JCAP {\bf 1909}, no. 09, 040 (2019)
  [arXiv:1906.10176 [hep-th]].

\bibitem{king}
  M.~C.~Romao and S.~F.~King,
  JHEP {\bf 1707}, 033 (2017)
  [arXiv:1703.08333 [hep-ph]];
  S.~F.~King and E.~Perdomo,
  JHEP {\bf 1905}, 211 (2019)
  [arXiv:1903.08448 [hep-ph]].

    \bibitem{KLno-scale}
R.~Kallosh and A.~Linde,
  JCAP {\bf 1306}, 028 (2013)
  [arXiv:1306.3214 [hep-th]].
 
\bibitem{FKR}
F.~Farakos, A.~Kehagias and A.~Riotto,
  Nucl.\ Phys.\ B {\bf 876}, 187 (2013)
  [arXiv:1307.1137 [hep-th]].
  
  \bibitem{FeKR}
  S.~Ferrara, A.~Kehagias and A.~Riotto,
  Fortsch.\ Phys.\  {\bf 62}, 573 (2014)
  [arXiv:1403.5531 [hep-th]];
S.~Ferrara, A.~Kehagias and A.~Riotto,
  Fortsch.\ Phys.\  {\bf 63}, 2 (2015)
  [arXiv:1405.2353 [hep-th]];
 R.~Kallosh, A.~Linde, B.~Vercnocke and W.~Chemissany,
  JCAP {\bf 1407}, 053 (2014)
  [arXiv:1403.7189 [hep-th]];
 K.~Hamaguchi, T.~Moroi and T.~Terada,
  Phys.\ Lett.\ B {\bf 733}, 305 (2014)
  [arXiv:1403.7521 [hep-ph]];
 J.~Ellis, M.~A.~G.~Garc\'ia, D.~V.~Nanopoulos and K.~A.~Olive,
  JCAP {\bf 1405}, 037 (2014)
  [arXiv:1403.7518 [hep-ph]];
J.~Ellis, M.~A.~G.~Garc{\' i}a, D.~V.~Nanopoulos and K.~A.~Olive,
  JCAP {\bf 1408}, 044 (2014)
  [arXiv:1405.0271 [hep-ph]].

\bibitem{adfs}
I.~Antoniadis, E.~Dudas, S.~Ferrara and A.~Sagnotti,
  Phys.\ Lett.\ B {\bf 733}, 32 (2014)
  [arXiv:1403.3269 [hep-th]].

 \bibitem{EGNO4} 
  J.~Ellis, M.~A.~G.~Garcia, D.~V.~Nanopoulos and K.~A.~Olive,
  JCAP {\bf 1510}, 10, 003 (2015)
  [arXiv:1503.08867 [hep-ph]].
  
  \bibitem{dgmo}
   E.~Dudas, T.~Gherghetta, Y.~Mambrini and K.~A.~Olive,
  Phys.\ Rev.\ D {\bf 96}, no. 11, 115032 (2017)
  [arXiv:1710.07341 [hep-ph]];


\bibitem{others}
  S.~Ferrara, R.~Kallosh, A.~Linde and M.~Porrati,
  Phys.\ Rev.\ D {\bf 88} (2013) 8,  085038
  [arXiv:1307.7696 [hep-th]];
  W.~Buchm\"uller, V.~Domcke and C.~Wieck,
  Phys.\ Lett.\ B {\bf 730}, 155 (2014)
  [arXiv:1309.3122 [hep-th]];
  C.~Pallis,
  JCAP {\bf 1404}, 024 (2014)
  [arXiv:1312.3623 [hep-ph]];
   C.~Pallis,
  JCAP {\bf 1408}, 057 (2014)
  [arXiv:1403.5486 [hep-ph]];
     W.~Buchmuller, E.~Dudas, L.~Heurtier and C.~Wieck,
  JHEP {\bf 1409}, 053 (2014)
  [arXiv:1407.0253 [hep-th]];
    J.~Ellis, M.~A.~G.~Garc\'ia, D.~V.~Nanopoulos and K.~A.~Olive,
  JCAP {\bf 1501}, no. 01, 010 (2015)
  [arXiv:1409.8197 [hep-ph]];
   T.~Terada, Y.~Watanabe, Y.~Yamada and J.~Yokoyama,
  JHEP {\bf 1502}, 105 (2015)
  [arXiv:1411.6746 [hep-ph]];
  W.~Buchmuller, E.~Dudas, L.~Heurtier, A.~Westphal, C.~Wieck and M.~W.~Winkler,
  JHEP {\bf 1504}, 058 (2015)
  [arXiv:1501.05812 [hep-th]];
 A.~B.~Lahanas and K.~Tamvakis,
  Phys.\ Rev.\ D {\bf 91}, no. 8, 085001 (2015)
  [arXiv:1501.06547 [hep-th]];
I.~Dalianis and F.~Farakos,
  JCAP {\bf 1507}, no. 07, 044 (2015)
  [arXiv:1502.01246 [gr-qc]].
  I.~Garg and S.~Mohanty,
  Phys.\ Lett.\ B {\bf 751}, 7 (2015)
  [arXiv:1504.07725 [hep-ph]];
  J.~Ellis, M.~A.~G.~Garc{\' i}a, D.~V.~Nanopoulos and K.~A.~Olive,
  JCAP {\bf 1507}, no. 07, 050 (2015)
  [arXiv:1505.06986 [hep-ph]];
  E.~Dudas and C.~Wieck,
  JHEP {\bf 1510}, 062 (2015)
  [arXiv:1506.01253 [hep-th]];
M.~Scalisi,
  JHEP {\bf 1512}, 134 (2015)
  [arXiv:1506.01368 [hep-th]];
  S.~Ferrara, A.~Kehagias and M.~Porrati,
  JHEP {\bf 1508}, 001 (2015)
  [arXiv:1506.01566 [hep-th]];
  J.~Ellis, M.~A.~G.~Garc{\' i}a, D.~V.~Nanopoulos and K.~A.~Olive,
  Class.\ Quant.\ Grav.\  {\bf 33}, no. 9, 094001 (2016)
  [arXiv:1507.02308 [hep-ph]];
  A.~Addazi and M.~Y.~Khlopov,
  Phys.\ Lett.\ B {\bf 766}, 17 (2017)
  [arXiv:1612.06417 [gr-qc]];
   C.~Pallis and N.~Toumbas,
 Adv.\ High Energy Phys.\  {\bf 2017}, 6759267 (2017)
  [arXiv:1612.09202 [hep-ph]];
  T.~Kobayashi, O.~Seto and T.~H.~Tatsuishi,
  PTEP {\bf 2017}, no. 12, 123B04 (2017)
  [arXiv:1703.09960 [hep-th]];
 I.~Garg and S.~Mohanty,
  Int.\ J.\ Mod.\ Phys.\ A {\bf 33}, no. 21, 1850127 (2018)
  [arXiv:1711.01979 [hep-ph]];
   W.~Ahmed and A.~Karozas,
  Phys.\ Rev.\ D {\bf 98}, no. 2, 023538 (2018)
  [arXiv:1804.04822 [hep-ph]].
   Y.~Cai, R.~Deen, B.~A.~Ovrut and A.~Purves,
  JHEP {\bf 1809}, 001 (2018)
  [arXiv:1804.07848 [hep-th]].
   S.~Khalil, A.~Moursy, A.~K.~Saha and A.~Sil,
  Phys.\ Rev.\ D {\bf 99}, no. 9, 095022 (2019)
  [arXiv:1810.06408 [hep-ph]].

 \bibitem{egnno1}
  J.~Ellis, M.~A.~G.~Garcia, N.~Nagata, D.~V.~Nanopoulos and K.~A.~Olive,
  JCAP {\bf 1611}, no. 11, 018 (2016)
  [arXiv:1609.05849 [hep-ph]].
  
  \bibitem{egnno23}
   J.~Ellis, M.~A.~G.~Garcia, N.~Nagata, D.~V.~Nanopoulos and K.~A.~Olive,
  JCAP {\bf 1707}, no. 07, 006 (2017)
  [arXiv:1704.07331 [hep-ph]];
  J.~Ellis, M.~A.~G.~Garcia, N.~Nagata, D.~V.~Nanopoulos and K.~A.~Olive,
  JCAP {\bf 1904}, no. 04, 009 (2019)
  [arXiv:1812.08184 [hep-ph]].
  
 \bibitem{egko}
S.~A.~R.~Ellis, T.~Gherghetta, K.~Kaneta and K.~A.~Olive,
  Phys.\ Rev.\ D {\bf 98}, no. 5, 055009 (2018)
  [arXiv:1807.06488 [hep-ph]].
  
    \bibitem{GN2}
  H.~Georgi and D.~V.~Nanopoulos,
  Nucl.\ Phys.\ B {\bf 159}, 16 (1979);
  C.~E.~Vayonakis,
  Phys.\ Lett.\ B {\bf 82}, 224 (1979)
  [Phys.\ Lett.\  {\bf 83B}, 421 (1979)];
  A.~Masiero,
  Phys.\ Lett.\ B {\bf 93}, 295 (1980);
  Q.~Shafi, M.~Sondermann and C.~Wetterich,
  Phys.\ Lett.\ B {\bf 92}, 304 (1980);
  F.~del Aguila and L.~E.~Ibanez,
  Nucl.\ Phys.\ B {\bf 177}, 60 (1981);
  R.~N.~Mohapatra and G.~Senjanovic,
  Phys.\ Rev.\ D {\bf 27}, 1601 (1983);
  M.~Fukugita and T.~Yanagida,
  In *Fukugita, M. (ed.), Suzuki, A. (ed.): Physics and astrophysics of neutrinos* 1-248. and Kyoto Univ. - YITP-K-1050 (93/12,rec.Feb.94) 248 p. C.
  
    \bibitem{moqz}
  Y.~Mambrini, K.~A.~Olive, J.~Quevillon and B.~Zaldivar,
  Phys.\ Rev.\ Lett.\  {\bf 110}, no. 24, 241306 (2013)
  [arXiv:1302.4438 [hep-ph]].
  
   
\bibitem{mnoqz}
Y.~Mambrini, N.~Nagata, K.~A.~Olive, J.~Quevillon and J.~Zheng,
  Phys.\ Rev.\ D {\bf 91}, no. 9, 095010 (2015)
  [arXiv:1502.06929 [hep-ph]].
  
  \bibitem{noz}
  N.~Nagata, K.~A.~Olive and J.~Zheng,
  JHEP {\bf 1510}, 193 (2015)
  [arXiv:1509.00809 [hep-ph]].
  
  \bibitem{mnoz}
   Y.~Mambrini, N.~Nagata, K.~A.~Olive and J.~Zheng,
  Phys.\ Rev.\ D {\bf 93}, no. 11, 111703 (2016)
  [arXiv:1602.05583 [hep-ph]].
  
    \bibitem{pp}
  H.~Pagels and J.~R.~Primack,
  Phys.\ Rev.\ Lett.\  {\bf 48}, 223 (1982).
  
  
  \bibitem{nos}
  D.~V.~Nanopoulos, K.~A.~Olive and M.~Srednicki,
  Phys.\ Lett.\ B {\bf 127}, 30 (1983).

  
  \bibitem{oss}
  K.~A.~Olive, D.~N.~Schramm and M.~Srednicki,
  Nucl.\ Phys.\ B {\bf 255}, 495 (1985).

\bibitem{eoss5}
J.~R.~Ellis, K.~A.~Olive, Y.~Santoso and V.~C.~Spanos,
  Phys.\ Lett.\ B {\bf 588}, 7 (2004)
  [hep-ph/0312262].
  
 


 \bibitem{0404231}
 J.~L.~Feng, S.~f.~Su and F.~Takayama,
  Phys.\ Rev.\ D {\bf 70}, 063514 (2004)
  [hep-ph/0404198];
 J.~L.~Feng, S.~Su and F.~Takayama,
  Phys.\ Rev.\ D {\bf 70} (2004) 075019
  [arXiv:hep-ph/0404231].
  
\bibitem{stef}
   F.~D.~Steffen,
  JCAP {\bf 0609}, 001 (2006)
  [hep-ph/0605306].
  
\bibitem{buch}
W.~Buchmuller, L.~Covi, K.~Hamaguchi, A.~Ibarra and T.~Yanagida,
  JHEP {\bf 0703}, 037 (2007)
  [hep-ph/0702184 [HEP-PH]];
  W.~Buchmuller,
  AIP Conf.\ Proc.\  {\bf 1200}, 155 (2010)
  [arXiv:0910.1870 [hep-ph]].
  
  \bibitem{rosz}
   S.~Bailly, K.~Y.~Choi, K.~Jedamzik and L.~Roszkowski,
  JHEP {\bf 0905}, 103 (2009)
  [arXiv:0903.3974 [hep-ph]].
  
  \bibitem{covi}
   L.~Covi, J.~Hasenkamp, S.~Pokorski and J.~Roberts,
  JHEP {\bf 0911}, 003 (2009)
  [arXiv:0908.3399 [hep-ph]].
  
  
    \bibitem{nosusy}
  M.~Aaboud {\it et al.} [ATLAS Collaboration],
  JHEP {\bf 1806}, 107 (2018)
  [arXiv:1711.01901 [hep-ex]];
  M.~Aaboud {\it et al.} [ATLAS Collaboration],
  Phys.\ Rev.\ D {\bf 97}, no. 11, 112001 (2018)
  [arXiv:1712.02332 [hep-ex]];
  A.~M.~Sirunyan {\it et al.} [CMS Collaboration],
  Eur.\ Phys.\ J.\ C {\bf 77}, no. 10, 710 (2017)
  [arXiv:1705.04650 [hep-ex]];
  A.~M.~Sirunyan {\it et al.} [CMS Collaboration],
  JHEP {\bf 1805}, 025 (2018)
  [arXiv:1802.02110 [hep-ex]].
  
  \bibitem{LUX}
  D.~S.~Akerib {\it et al.} [LUX Collaboration],
  Phys.\ Rev.\ Lett.\  {\bf 118} (2017) no.2,  021303
  [arXiv:1608.07648 [astro-ph.CO]].

  \bibitem{PANDAX}
  X.~Cui {\it et al.} [PandaX-II Collaboration],
  Phys.\ Rev.\ Lett.\  {\bf 119} (2017) no.18,  181302
  [arXiv:1708.06917 [astro-ph.CO]].

  \bibitem{XENON}
 E.~Aprile {\it et al.} [XENON Collaboration],
  Phys.\ Rev.\ Lett.\  {\bf 121} (2018) no.11,  111302
  [arXiv:1805.12562 [astro-ph.CO]].
  
  \bibitem{cmssm}
  J.~Ellis, J.~L.~Evans, F.~Luo, N.~Nagata, K.~A.~Olive and P.~Sandick,
  Eur.\ Phys.\ J.\ C {\bf 76}, no. 1, 8 (2016)
  [arXiv:1509.08838 [hep-ph]];
   J.~Ellis, J.~L.~Evans, A.~Mustafayev, N.~Nagata and K.~A.~Olive,
  Eur.\ Phys.\ J.\ C {\bf 76}, no. 11, 592 (2016)
  [arXiv:1608.05370 [hep-ph]];
  J.~Ellis, J.~L.~Evans, N.~Nagata, D.~V.~Nanopoulos and K.~A.~Olive,
  Eur.\ Phys.\ J.\ C {\bf 77}, no. 4, 232 (2017)
  [arXiv:1702.00379 [hep-ph]];
  J.~Ellis, J.~L.~Evans, F.~Luo, K.~A.~Olive and J.~Zheng,
  Eur.\ Phys.\ J.\ C {\bf 78}, no. 5, 425 (2018)
  [arXiv:1801.09855 [hep-ph]];
   E.~Bagnaschi {\it et al.},
  Eur.\ Phys.\ J.\ C {\bf 79}, no. 2, 149 (2019)
  [arXiv:1810.10905 [hep-ph]].


\bibitem{pgm}
  M.~Ibe, T.~Moroi and T.~T.~Yanagida,
  Phys.\ Lett.\ B {\bf 644}, 355 (2007)
  [hep-ph/0610277];
 M.~Ibe and T.~T.~Yanagida,
  Phys.\ Lett.\ B {\bf 709}, 374 (2012)
  [arXiv:1112.2462 [hep-ph]];
 M.~Ibe, S.~Matsumoto and T.~T.~Yanagida,
  Phys.\ Rev.\ D {\bf 85}, 095011 (2012)
  [arXiv:1202.2253 [hep-ph]];
  J.~L.~Evans and K.~A.~Olive,
  Phys.\ Rev.\ D {\bf 90}, no. 11, 115020 (2014)
  [arXiv:1408.5102 [hep-ph]];
    J.~L.~Evans, M.~Ibe, K.~A.~Olive and T.~T.~Yanagida,
  Phys.\ Rev.\ D {\bf 91}, 055008 (2015)
  [arXiv:1412.3403 [hep-ph]];
J.~L.~Evans, N.~Nagata and K.~A.~Olive,
  arXiv:1902.09084 [hep-ph].

  
  \bibitem{eioy}
  J.~L.~Evans, M.~Ibe, K.~A.~Olive and T.~T.~Yanagida,
  Eur.\ Phys.\ J.\ C {\bf 73}, 2468 (2013)
  [arXiv:1302.5346 [hep-ph]];
J.~L.~Evans, K.~A.~Olive, M.~Ibe and T.~T.~Yanagida,
  Eur.\ Phys.\ J.\ C {\bf 73}, no. 10, 2611 (2013)
  [arXiv:1305.7461 [hep-ph]].
  
 \bibitem{hssusy}
  G.~F.~Giudice and A.~Strumia,
  Nucl.\ Phys.\ B {\bf 858}, 63 (2012)
  [arXiv:1108.6077 [hep-ph]];
    E.~Bagnaschi, G.~F.~Giudice, P.~Slavich and A.~Strumia,
  JHEP {\bf 1409}, 092 (2014)
  [arXiv:1407.4081 [hep-ph]].
  
  
  \bibitem{eev}
  E.~Dudas, Y.~Mambrini and K.~Olive,
  Phys.\ Rev.\ Lett.\  {\bf 119}, no. 5, 051801 (2017)
  [arXiv:1704.03008 [hep-ph]].
  

  
\bibitem{dgkmo}  
   E.~Dudas, T.~Gherghetta, K.~Kaneta, Y.~Mambrini and K.~A.~Olive,
  Phys.\ Rev.\ D {\bf 98}, no. 1, 015030 (2018)
  [arXiv:1805.07342 [hep-ph]].
  
  \bibitem{kmo}
  K.~Kaneta, Y.~Mambrini and K.~A.~Olive,
  Phys.\ Rev.\ D {\bf 99}, no. 6, 063508 (2019)
  [arXiv:1901.04449 [hep-ph]].
  
  \bibitem{ekn}
  J.~R.~Ellis, J.~E.~Kim and D.~V.~Nanopoulos,
  Phys.\ Lett.\ B {\bf 145}, 181 (1984).
  
    \bibitem{Moroi:1995fs} 
  T.~Moroi,
  hep-ph/9503210.
  
  \bibitem{enor}
  J.~R.~Ellis, D.~V.~Nanopoulos, K.~A.~Olive and S.~J.~Rey,
  Astropart.\ Phys.\  {\bf 4}, 371 (1996)
  [hep-ph/9505438].
  
      \bibitem{Giudice:1999am} 
  G.~F.~Giudice, A.~Riotto and I.~Tkachev,
  JHEP {\bf 9911}, 036 (1999)
  [hep-ph/9911302].


  
   \bibitem{bbb}
   M.~Bolz, A.~Brandenburg and W.~Buchmuller,
  Nucl.\ Phys.\ B {\bf 606}, 518 (2001)
  [Erratum-ibid.\ B {\bf 790}, 336 (2008)]
  [hep-ph/0012052];
    J.~Pradler and F.~D.~Steffen,
  Phys.\ Rev.\ D {\bf 75}, 023509 (2007)
  [hep-ph/0608344];
 V.~S.~Rychkov and A.~Strumia,
  Phys.\ Rev.\ D {\bf 75}, 075011 (2007)
  [hep-ph/0701104].
  
  \bibitem{egnop}
  J.~Ellis, M.~A.~G.~Garcia, D.~V.~Nanopoulos, K.~A.~Olive and M.~Peloso,
  JCAP {\bf 1603}, no. 03, 008 (2016)
  [arXiv:1512.05701 [astro-ph.CO]].
  
  \bibitem{bcdm}
  K.~Benakli, Y.~Chen, E.~Dudas and Y.~Mambrini,
Phys.\ Rev.\ D {\bf 95}, no. 9, 095002 (2017)
  [arXiv:1701.06574 [hep-ph]].
  
  \bibitem{gmop}
  M.~A.~G.~Garcia, Y.~Mambrini, K.~A.~Olive and M.~Peloso,
  Phys.\ Rev.\ D {\bf 96}, no. 10, 103510 (2017)
  [arXiv:1709.01549 [hep-ph]].

  
            \bibitem{so10dm}
   M.~Kadastik, K.~Kannike and M.~Raidal,
  Phys.\ Rev.\ D {\bf 80} (2009) 085020
   [Erratum-ibid.\ D {\bf 81} (2010) 029903]
  [arXiv:0907.1894 [hep-ph]];
  M.~Kadastik, K.~Kannike and M.~Raidal,
  Phys.\ Rev.\ D {\bf 81}, 015002 (2010)
  [arXiv:0903.2475 [hep-ph]];
  M.~Frigerio and T.~Hambye,
  Phys.\ Rev.\ D {\bf 81} (2010) 075002
  [arXiv:0912.1545 [hep-ph]].

\bibitem{enoz}
J.~L.~Evans, N.~Nagata, K.~A.~Olive and J.~Zheng,
  JHEP {\bf 1602}, 120 (2016)
  [arXiv:1512.02184 [hep-ph]].
  
  
  
    \bibitem{FY}
M.~Fukugita and T.~Yanagida,
  Phys.\ Lett.\ B {\bf 174} (1986) 45.


\bibitem{leptogenesis}
 S.~Davidson and A.~Ibarra,
  Phys.\ Lett.\ B {\bf 535} (2002) 25
  [hep-ph/0202239];
 E.~Nardi, Y.~Nir, E.~Roulet and J.~Racker,
  JHEP {\bf 0601} (2006) 164
  [hep-ph/0601084];
  A.~Abada, S.~Davidson, A.~Ibarra, F.~-X.~Josse-Michaux, M.~Losada and A.~Riotto,
  JHEP {\bf 0609} (2006) 010
  [hep-ph/0605281];
 R.~Barbieri, P.~Creminelli, A.~Strumia and N.~Tetradis,
  Nucl.\ Phys.\ B {\bf 575} (2000) 61
  [hep-ph/9911315];
  M.~Raidal, A.~Strumia and K.~Turzynski,
  Phys.\ Lett.\ B {\bf 609} (2005) 351
   [Erratum-ibid.\ B {\bf 632} (2006) 752]
  [hep-ph/0408015];
 A.~Pilaftsis and T.~E.~J.~Underwood,
  Nucl.\ Phys.\ B {\bf 692} (2004) 303
  [hep-ph/0309342];
P.~S.~Bhupal Dev, P.~Millington, A.~Pilaftsis and D.~Teresi,
  Nucl.\ Phys.\ B {\bf 886} (2014) 569
  [arXiv:1404.1003 [hep-ph]].


  
  \bibitem{seesaw}
P.~Minkowski,
  Phys.\ Lett.\ B {\bf 67} (1977) 421;
M.~Gell-Mann, P.~Ramond and R.~Slansky, in {\it Supergravity}, eds. D. Freedman and P. Van Nieuwenhuizen
(North Holland, Amsterdam, 1979), pp. 315-321. ISBN 044485438x;
T. Yanagida, in {\it Proceedings of the Workshop on 
the Unified Theory and The Baryon Number of the Universe}, eds O. Sawada  
and S. Sugamoto. KEK79-18 (1979);
R.~N.~Mohapatra and G.~Senjanovic,
  Phys.\ Rev.\ Lett.\  {\bf 44}, 912 (1980);
J.~Schechter and J.~W.~F.~Valle,
  Phys.\ Rev.\ D {\bf 22} (1980) 2227;
J.~Schechter and J.~W.~F.~Valle,
  Phys.\ Rev.\ D {\bf 25} (1982) 774.
  
  
  \bibitem{spha1}
  V.~A.~Kuzmin, V.~A.~Rubakov and M.~E.~Shaposhnikov,
Phys.\ Lett.\  {\bf 155B}, 36 (1985)..

\bibitem{spha2}
  S.~Y.~Khlebnikov and M.~E.~Shaposhnikov,
Nucl.\ Phys.\ B {\bf 308}, 885 (1988)..
  J.~A.~Harvey and M.~S.~Turner,
Phys.\ Rev.\ D {\bf 42}, 3344 (1990)..
  
      \bibitem{Cecotti}
S.~Cecotti,
  Phys.\ Lett.\ B {\bf 190} (1987) 86.

  
      \bibitem{EKN3}
   J.~R.~Ellis, C.~Kounnas and D.~V.~Nanopoulos,
  Phys.\ Lett.\ B {\bf 143}, 410 (1984).
  
  
    \bibitem{pol}
J. Polonyi, Budapest preprint KFKI-1977-93 (1977).

\bibitem{enno}
 J.~Ellis, B.~Nagaraj, D.~V.~Nanopoulos and K.~A.~Olive,
  JHEP {\bf 1811}, 110 (2018)
  [arXiv:1809.10114 [hep-th]];
  J.~Ellis, B.~Nagaraj, D.~V.~Nanopoulos, K.~A.~Olive and S.~Verner,
  JHEP {\bf 1910}, 161 (2019)
  doi:10.1007/JHEP10(2019)161
  [arXiv:1907.09123 [hep-th]].

   \bibitem{dine}
  M.~Dine, R.~Kitano, A.~Morisse and Y.~Shirman,
  Phys.\ Rev.\ D {\bf 73}, 123518 (2006)
  [hep-ph/0604140];
  R.~Kitano,
  Phys.\ Lett.\ B {\bf 641}, 203 (2006)
  [hep-ph/0607090];
  R.~Kallosh and A.~D.~Linde,
  JHEP {\bf 0702}, 002 (2007)
  [hep-th/0611183];  
H.~Abe, T.~Higaki and T.~Kobayashi,
  Phys.\ Rev.\ D {\bf 76} (2007) 105003
  [arXiv:0707.2671 [hep-th]];
J.~Fan, M.~Reece and L.-T. Wang,
JHEP {\bf 1109}, 126 (2011) [arXiv:1106.6044 [hep-ph]].

  \bibitem{Dudas:2006gr}
  E.~Dudas, C.~Papineau and S.~Pokorski,
  JHEP {\bf 0702}, 028 (2007)
  [hep-th/0610297];
  H.~Abe, T.~Higaki, T.~Kobayashi and Y.~Omura,
  Phys.\ Rev.\ D {\bf 75}, 025019 (2007)
  [hep-th/0611024].
  
   \bibitem{klor}
  R.~Kallosh, A.~Linde, K.~A.~Olive and T.~Rube,
  Phys.\ Rev.\ D {\bf 84}, 083519 (2011)
  [arXiv:1106.6025 [hep-th]];
   A.~Linde, Y.~Mambrini and K.~A.~Olive,
  Phys.\ Rev.\ D {\bf 85}, 066005 (2012)
  [arXiv:1111.1465 [hep-th]].

  \bibitem{dlmmo}
   E.~Dudas, A.~Linde, Y.~Mambrini, A.~Mustafayev and K.~A.~Olive,
  Eur.\ Phys.\ J.\ C {\bf 73}, no. 1, 2268 (2013)
  [arXiv:1209.0499 [hep-ph]].
  
  \bibitem{nataya}
K.~Nakayama, F.~Takahashi and T.~T.~Yanagida,
  Phys.\ Lett.\ B {\bf 718}, 526 (2012)
  [arXiv:1209.2583 [hep-ph]].
  
  \bibitem{ADinf}
 M.~A.~G.~Garcia and K.~A.~Olive,
  JCAP {\bf 1309}, 007 (2013)
  [arXiv:1306.6119 [hep-ph]].

  
  \bibitem{ego} 
  J.~L.~Evans, M.~A.~G.~Garcia and K.~A.~Olive,
  JCAP {\bf 1403}, 022 (2014)
  [arXiv:1311.0052 [hep-ph]].
  
  \bibitem{lept2}
  P.~H.~Frampton, S.~L.~Glashow and T.~Yanagida,
Phys.\ Lett.\ B {\bf 548}, 119 (2002).
[hep-ph/0208157].



\bibitem{pdg}
  M.~Tanabashi {\it et al.} [Particle Data Group],
Phys.\ Rev.\ D {\bf 98}, no. 3, 030001 (2018)..

  \bibitem{luty}
  M.~A.~Luty,
  Phys.\ Rev.\ D {\bf 45}, 455 (1992).


\bibitem{CPviol}
  L.~Covi, E.~Roulet and F.~Vissani,
Phys.\ Lett.\ B {\bf 384}, 169 (1996).
[hep-ph/9605319].
  M.~Flanz, E.~A.~Paschos and U.~Sarkar,
Phys.\ Lett.\ B {\bf 345}, 248 (1995), Erratum: [Phys.\ Lett.\ B {\bf 384}, 487 (1996)], Erratum: [Phys.\ Lett.\ B {\bf 382}, 447 (1996)].
[hep-ph/9411366].

\bibitem{cdo}
 B.~A.~Campbell, S.~Davidson and K.~A.~Olive,
  Phys.\ Lett.\ B {\bf 303}, 63 (1993)
  [hep-ph/9302222];
B.~A.~Campbell, S.~Davidson and K.~A.~Olive,
  Nucl.\ Phys.\ B {\bf 399}, 111 (1993)
  [hep-ph/9302223].


\bibitem{giudicereh}
  G.~F.~Giudice, M.~Peloso, A.~Riotto and I.~Tkachev,
  JHEP {\bf 9908} (1999) 014
  doi:10.1088/1126-6708/1999/08/014
  [hep-ph/9905242].

\bibitem{leptinf1}
  T.~Asaka, K.~Hamaguchi, M.~Kawasaki and T.~Yanagida,
Phys.\ Lett.\ B {\bf 464}, 12 (1999).
[hep-ph/9906366].

\bibitem{leptinf2}
  T.~Asaka, K.~Hamaguchi, M.~Kawasaki and T.~Yanagida,
Phys.\ Rev.\ D {\bf 61}, 083512 (2000).
[hep-ph/9907559].

\bibitem{egnno5}
J.~Ellis, M.~A.~G.~Garcia, N.~Nagata, D.~V.~Nanopoulos and K.~A.~Olive,
  Phys.\ Lett.\ B {\bf 797}, 134864 (2019)
  [arXiv:1906.08483 [hep-ph]];
J.~Ellis, M.~A.~G.~Garcia, N.~Nagata, D.~V.~Nanopoulos and K.~A.~Olive,
  arXiv:1910.11755 [hep-ph].



  
 \bibitem{dgkmo2}
  E.~Dudas, T.~Gherghetta, K.~Kaneta, Y.~Mambrini and K.~A.~Olive,
  Phys.\ Rev.\ D {\bf 100}, no. 3, 035004 (2019)
  doi:10.1103/PhysRevD.100.035004
  [arXiv:1905.09243 [hep-ph]].


\bibitem{Ferrara:2016ntj} 
  S.~Ferrara and A.~Van Proeyen,
  Fortsch.\ Phys.\  {\bf 64}, no. 11-12, 896 (2016)
  doi:10.1002/prop.201600100
  [arXiv:1609.08480 [hep-th]].
  
  \bibitem{nop}
  H.~P.~Nilles, K.~A.~Olive and M.~Peloso,
  Phys.\ Lett.\ B {\bf 522}, 304 (2001)
  doi:10.1016/S0370-2693(01)01300-4
  [hep-ph/0107212].
  
  \bibitem{IC}
M.~G.~Aartsen {\it et al.} [IceCube Collaboration],
  Eur.\ Phys.\ J.\ C {\bf 78} (2018) no.10,  831
  [arXiv:1804.03848 [astro-ph.HE]];
C.~Rott,
  PoS ICRC {\bf 2017} (2017) 1119
  [arXiv:1712.00666 [astro-ph.HE]];
  J.~Stettner and H.~Dujmovic,
  PoS ICRC {\bf 2017}, 923 (2018).




  
\bibitem{ekoty}
M.~Endo, K.~Kadota, K.~A.~Olive, F.~Takahashi and T.~T.~Yanagida,
  JCAP {\bf 0702}, 018 (2007)
  doi:10.1088/1475-7516/2007/02/018
  [hep-ph/0612263].

 \bibitem{thermlept}
  G.~F.~Giudice, A.~Notari, M.~Raidal, A.~Riotto and A.~Strumia,
Nucl.\ Phys.\ B {\bf 685}, 89 (2004).
[hep-ph/0310123].

\end{thebibliography}
\end{document}